\documentclass[11pt, a4paper]{article}
\usepackage[affil-it]{authblk}

\usepackage{fancyhdr}
\fancypagestyle{firstpage}{
  \fancyhf{}  
   
 \fancyhead[R]{KUNS-3059,~UT-Komaba/25-6}
}
\fancypagestyle{nonfirstpage}{
  \fancyhf{}  
   
 \fancyfoot[C]{\hyperlink{contents}{\thepage}} 
}

\usepackage{dcolumn}

\usepackage[utf8]{inputenc}

\usepackage{bm}
\usepackage{cite}
\usepackage{comment} 
\usepackage{mathtools}
\usepackage[top=30truemm,bottom=30truemm,left=21truemm,right=21truemm]{geometry}
\usepackage{graphicx}
\usepackage{amsmath,latexsym,amssymb,mathrsfs,ascmac}
\usepackage{multirow}
\usepackage{braket}
\usepackage{physics}

\usepackage{color}

\usepackage[colorlinks=true,urlcolor=blue,anchorcolor=black,citecolor=blue,linkcolor=black,filecolor=black,menucolor=black,linktocpage=true,pdfproducer=medialab,pdfa=true,pdftex]{hyperref}

\usepackage{enumerate}
\usepackage{tensor}
\usepackage{cleveref}
\newcommand{\nn}{\nonumber \\}
\newcommand{\ep}{\epsilon}
\newcommand{\epv}{\varepsilon}
\newcommand{\phiv}{\varphi}

\newcommand{\del}{\partial}

\newcommand{\mr}[1]{\mathrm{#1}}
\newcommand{\mc}[1]{\mathcal{#1}}

\newcommand{\lie}[1]{\mathsterling_{#1}}
\newcommand{\tens}{\tensor}
\newcommand{\bg}[1]{\stackrel{\circ}{#1} \!{\!}}

\usepackage{setspace}

 \makeatletter
 
 \@addtoreset{equation}{section}
 \makeatother
 
\begin{document}
\newgeometry{top=30truemm,bottom=30truemm,left=21truemm,right=21truemm}

\title{\bfseries \fontsize{14pt}{18pt}\selectfont
Apparent Horizons Associated with Dynamical Black Hole Entropy
\vspace{1em}
}

\date{}

\renewcommand\Authfont{\normalsize\normalfont\raggedright\bfseries}
\renewcommand\Affilfont{\normalfont\itshape\raggedright\fontsize{10.4pt}{12}\selectfont}
\renewcommand\Authand{ and }
\renewcommand\Authands{, and }
\newcommand{\email}[1]{\thanks{\href{mailto:#1}{\texttt{#1}}}}

\author[1]{Hideo Furugori\email{h-furugori@gauge.scphys.kyoto-u.ac.jp}}
\affil[1]{Department of Physics, Kyoto University, Kyoto 606-8502, Japan}

\author[2]{Kanji Nishii\email{kanji.nishii@gmail.com}}
\affil[2]{Graduate School of Arts and Sciences, University of Tokyo, Komaba, Meguro-ku, Tokyo 153-8902, Japan}

\author[3]{Daisuke Yoshida\email{dyoshida@math.nagoya-u.ac.jp}}
\affil[3]{Department of Mathematics, Nagoya University, Nagoya 464-8602, Japan}

\author[2]{Kaho Yoshimura\email{yoshimura-kaho848@g.ecc.u-tokyo.ac.jp}}

\maketitle
\thispagestyle{firstpage}

\noindent
{\fontsize{11.2pt}{12}\selectfont
{\bfseries\large Abstract:}
We define entropic marginally outer trapped surfaces (E-MOTSs), associated with a given entropy density, as a generalization of apparent horizons. We then show that, under first-order perturbations around a stationary black hole, the dynamical black hole entropy proposed by Hollands, Wald, and Zhang, defined on a background Killing horizon, can be expressed as the Wall entropy evaluated on an E-MOTS associated with it in any diffeomorphism invariant theory of gravity. Our result ensures that the Hollands--Wald--Zhang entropy reduces to the standard Wald entropy in each stationary regime of a dynamical black hole, thereby reinforcing the robustness of the dynamical entropy formulation.
}

\restoregeometry

\newpage

\pagestyle{nonfirstpage} 

\noindent\rule{\linewidth}{0.4pt}
\hypertarget{contents}{\tableofcontents}
\noindent\rule{\linewidth}{0.4pt}

\setstretch{1.2}

\section{Introduction}
Black hole thermodynamics provides insight into the quantum aspects of black holes. Among various thermodynamic quantities, entropy plays a particularly important role and has been discussed from many perspectives. Bekenstein and Hawking proposed that, for black holes in general relativity, the entropy is proportional to the area of the event horizon $A_{\text{event}}$ \cite{Bekenstein:1973ur,Hawking:1975vcx}, given by
\begin{align}
S_{\rm{BH}} = \frac{A_{\text{event}}}{4G\hbar}\,.
\end{align}
With this identification, the area increase law of the event horizon~\cite{Hawking:1971tu, Hawking:1971vc} is regarded as the second law of thermodynamics. Furthermore, this entropy is known to satisfy the first law of thermodynamics under stationary perturbations
on an arbitrary event horizon cross-section \cite{Bardeen:1973gs}. 
The event horizon is a global concept as it can be defined only after the whole future of the spacetime is fixed. In relation to this, when energy is injected in the future, the area increase law implies that the area already increases even during the stationary phase before the injection. This behavior is incompatible with the idea of identifying a locally stationary black hole with a state of thermal equilibrium, in which thermodynamic quantities remain constant. This perspective motivates a quasi-local redefinition of black hole entropy, and
various quasi-local notions have been introduced so far.

The event horizon for a stationary black hole coincides with an apparent horizon (i.e., a marginally outer trapped surface) from the viewpoint of kinematics of null congruences. From this perspective, various proposals have been developed to define black hole entropy in terms of the apparent horizon, for instance, the trapping horizon \cite{Hayward:1993wb,Hayward:1994bu,Hayward:1997jp}, the isolated horizon \cite{Ashtekar:1998sp}, and the dynamical horizon \cite{Ashtekar:2002ag,Ashtekar:2003hk}. As a general notion, one may consider hypersurfaces foliated by codimension-2 surfaces satisfying certain conditions, commonly referred to as tubes \cite{Ashtekar:2004cn,Ashtekar:2005ez,Booth:2005ng}. The area law along the tube is also discussed~\cite{Bousso:2015mqa,Mars:2024nyl}. 
However, the tube formed by apparent horizons can, in general, extend in spatial directions, e.g., when positive energy is injected into the black hole. In such cases, the area increase of the apparent horizon cannot be regarded as an increase along a time evolution, which undermines the analogy with thermodynamic laws.

On the other hand, the event horizon of a stationary black hole can also be regarded, from the viewpoint of symmetry, as the Killing horizon, where the Killing vector field becomes null.
From this perspective, Wald \cite{Wald:1993nt} proposed the entropy, known as the Wald entropy, as the Noether charge associated with diffeomorphism invariance, using the covariant phase space formalism in asymptotically flat spacetimes. This entropy depends on the gravitational action and reproduces the Bekenstein--Hawking's areal entropy for the Einstein--Hilbert action. 
Importantly, it is originally defined on the bifurcation surface $\mathcal{B}$ of the background stationary spacetime to ensure that the first law of thermodynamics holds under both stationary and non-stationary perturbations. 

Since the bifurcation surface is a special surface in spacetime, no time evolution, such as the second law of thermodynamics, can be argued. From this viewpoint, it is meaningful to extend the definition of entropy from the bifurcation surface to an arbitrary cross-section of the background Killing horizon. Iyer and Wald \cite{Iyer:1994ys} extended the definition of the Wald entropy to an arbitrary cross-section of the background Killing horizon. Throughout this paper, we refer to this generalized notion as the Iyer--Wald entropy, as we will define Eq.~\eqref{def IW entropy} in Sec.~\ref{subsec: HWZ}\footnote{
Our use of the term ``Iyer--Wald entropy" refers to the entropy $S$ defined in Eq.~(103) of Ref.~\cite{Iyer:1994ys}. Note, however, that another entropy formula, $S_{\text{dyn}}$ defined in Eq.~(104) of the same reference, is also referred to as the Iyer--Wald entropy in some literature, e.g., Ref.~\cite{Visser:2024pwz}. 
\label{fn: IY}
}.
As an alternative approach, Wall proposed another definition of entropy~\cite{Wall:2015raa,Wall:2024lbd} (the so-called Wall entropy) on arbitrary cross-sections, designed to satisfy the linearized second law under the condition that the system eventually settles into a stationary black hole. Both the Iyer--Wald entropy and the Wall entropy reduce to the Wald entropy when evaluated on the bifurcation surface, indicating that they satisfy the first law for arbitrary perturbations on that surface.
However, when they are evaluated on cross-sections other than the bifurcation surface, the corresponding first laws have not been established.

In this context, Hollands, Wald, and Zhang proposed the entropy of dynamical black holes based on the covariant phase space formalism \cite{Hollands:2024vbe}, which can apply to arbitrary diffeomorphism invariant gravitational actions. This entropy is defined on arbitrary cross-sections of the background Killing horizon and satisfies the first law of thermodynamics even for non-stationary perturbations, both of which are the desired properties. Importantly, for the Einstein--Hilbert action, the dynamical black hole entropy is proportional to the area of the apparent horizon $A_{\rm{app}}$,
\begin{align}
    S_{\rm{HWZ}}
    =\frac{A_{\text{app}}}{4G\hbar} \qquad (\text{for the Einstein--Hilbert action})\,, \label{SHWZ = Aapp}
\end{align}
even though the apparent horizon is not located on the background Killing horizon where the Hollands--Wald--Zhang entropy is evaluated. This agrees with the aforementioned ideas of defining the entropy based on the kinematics of null congruences. Moreover, the entropy is now explicitly evolving along the ``time" direction of the null hypersurface. In particular, Eq.~\eqref{SHWZ = Aapp} ensures that, in a stationary phase, the Hollands--Wald--Zhang entropy reduces to the Bekenstein--Hawking entropy that would apply if the spacetime were stationary eternally.
The purpose of this paper is to confirm that the same property holds for any diffeomorphism invariant gravitational action, not limited to the Einstein--Hilbert action, which is also conjectured in Ref.~\cite{Yan:2024gbz}.

Along this line, an extension of this property to $f(R)$ gravity was investigated in Ref.~\cite{Kong:2024sqc}.
There, the dynamical black hole entropy is found to be equivalent with the Iyer--Wald entropy evaluated on the surface where a generalization of the expansion in $f(R)$ gravity \cite{Matsuda:2020yvl} vanishes.
However, that analysis relies on a property specific to $f(R)$ gravity, namely the existence of the Einstein frame. In contrast, our analysis provides a proof using a method independent of specific details of the gravitational actions, which is a generalization of the method developed in Ref.~\cite{Visser:2024pwz} for the Einstein--Hilbert case.

Motivated by these developments, in this paper, we propose a definition of the \textit{entropic expansion} for any given entropy density. Based on it, we provide a definition of an \textit{Entropic Marginally Outer Trapped Surface (E-MOTS)}, as a generalization of the apparent horizon. Furthermore, we prove that the Hollands--Wald--Zhang entropy evaluated on a null hypersurface coincides with the Wall entropy evaluated on an E-MOTS.

This paper is organized as follows. In Sec.~\ref{sec: setup}, we introduce the affinely parametrized Gaussian null coordinates and the notion of boost weight as preliminaries. Sec.~\ref{sec: review on HWZ} reviews the concept of dynamical black hole entropy derived from non-stationary perturbations around a stationary black hole background, based on Refs.~\cite{Hollands:2024vbe,Visser:2024pwz}.
Sec.~\ref{sec EMOTS} is the main part of this paper. There, we provide the definitions of the entropic expansion and the entropic marginally outer trapped surface for any given entropy density. In addition, we prove a relation between the entropy evaluated on the E-MOTS and that on the background Killing horizon. We, then, apply our result to the Hollands--Wald--Zhang entropy and confirm that the Hollands--Wald--Zhang entropy is equivalent to the Wall entropy evaluated on an E-MOTS, which is our main result. Finally, we conclude with a summary and an outlook.

In this paper, lowercase latin indices such as $a,b,c,\cdots$ represent abstract indices following the conventions of the textbook~\cite{Wald:1984rg}. We work in a $\mathrm{D}$-dimensional spacetime. We adopt units in which the Boltzmann constant and the speed of light are set to unity, $k_B = c = 1$, while keeping the gravitational constant $G$\footnote{
By the gravitational constant $G$, we mean that the constant appearing in the Einstein--Hilbert Lagrangian as $L = R/16 \pi G$. In $\mr{D}$-dimensional spacetime, the gravitational constant $G$ can be expressed by the Newton constant $G_{\mr{N}}$ as $G = \frac{\Omega_{\mr{D}-2} (\mr{D} - 2)}{8 \pi (\mr{D} - 3)} G_{\mr{N}}$, where $\Omega_{\mr{D}-2} \coloneqq 2 \pi^{\frac{\mr{D}-1}{2}}/\Gamma\left(\frac{\mr{D}-1}{2}\right)$ is the area of a unit $(\mr{D} - 2)$-sphere. 
}
and the Planck constant $\hbar$ explicit.

\section{Basics of the Gaussian Null Coordinates\label{sec: setup}}
In this section, we provide a detailed review of the basic properties of affinely parametrized Gaussian null coordinates and the concept of boost weight. These notions play an important role in the technical developments of this paper, serving as useful tools for formulating and analyzing the geometric structures underlying our main results. By establishing this foundation, we aim to clarify how affinely parametrized Gaussian null coordinates and boost weight naturally emerge in the context of null hypersurfaces and facilitate a precise treatment of the dynamical entropy in later sections.
 
\subsection{The Affinely Parametrized Gaussian Null Coordinates\label{subsec AGNCs}}
Let us consider a $\mr{D}$-dimensional Lorentzian manifold $\mc{M}$ with a metric $g_{ab}$.
We introduce affinely parametrized Gaussian null coordinates~\cite{Penrose:1972ui, Hollands:2006rj,Hollands:2012sf}, $x^{\mu} = (u, v, \varphi^A)$ with $\varphi^A=(\varphi^1,\varphi^2,\cdots,\varphi^{\mathrm{D}-2})$ associated with a null hypersurface $\mathcal{N}$. That is, $v$ is chosen as an affine parameter along the null generators and the metric in this coordinate system is given by:
\begin{align}\label{metric on GNCs}
   & g_{ab}(u,v,\varphi^A) = - 2 (dv)_{(a} \left[ (du)_{b)} +\frac{u^2}{2} \alpha(u,v,\varphi^A) (dv)_{b)} + u \beta_{b)} (u,v,\varphi^A) \right]+ \gamma_{ab}(u,v,\varphi^A)\,,\\ \notag 
   & \beta_a(u,v,\varphi^A) = \beta_A (u,v,\varphi^A) (d\varphi^A)_a \,,\quad \gamma_{ab}(u,v,\varphi^A) = \gamma_{AB} (u,v,\varphi^A) (d \varphi^A)_{a} (d \varphi^B)_{b}\,,
\end{align}
where the $u = 0$ hypersurface corresponds to $\mathcal{N}$, and functions $\alpha, \beta_{A},$ and $\gamma_{AB}$ are assumed to be finite at $u = 0$. The inverse metric is evaluated as 
\begin{align}
    g^{ab}(u,v,\varphi^A) &=- 2 (\partial_u)^{(a} \left[ (\partial_v)^{b)} -\frac{u^2}{2}  \left(\alpha(u,v,\varphi^A) +\beta^2 (u,v,\varphi^A) \right) (\partial_u)^{b)}+u \beta^{b)}(u,v,\varphi^A) \right]
    \notag \\ & \quad  + \gamma^{ab}(u,v,\varphi^A)\,,
\label{inv metric on GNCs}
\end{align}
where $\gamma^{ab}$ is the inverse of $\gamma_{ab}$ and $\beta^2$ is given by $\beta^2=\beta_a \beta^a =\beta_A \beta^A$.

First, let us define one of the basis vectors
\begin{align}
    k^a \coloneqq (\partial _v)^a,
    \label{affine null k}
\end{align}
which is a future directed tangent vector of $u$-constant hypersurfaces and satisfies $k^a \gamma_{ab}=0 \,,\, k^a \beta_a  =0$.
The dual covector of $k^a$ is given by
\begin{align}
    k_a = g_{ab}k^b =- (du)_{a} -u^2 \alpha  (dv)_{a} -u \beta_a 
    \stackrel{\mathcal{N}}{=}-(du)_a \,,
    \label{k as normal}
\end{align}
and we can see that $k_{a}$ is the normal vector of $\mathcal{N}$.
Here, ``$\stackrel{\mathcal{N}}{=}$" means the equality evaluated on $\mathcal{N}$.
Then, we obtain
\begin{align}
    &k^a k_a = -u^2 \alpha \stackrel{\mathcal{N}}{=} 0 \label{k is null only on N}\,,\\
    &k^b \nabla_bk_a = -\Gamma^u_{va} -u\left[\partial_v \beta_A (d\varphi^A)_a +u \partial_v \alpha (dv)_a +u\alpha \Gamma^v_{va}+\beta_A \Gamma^A_{va}\right]
    \stackrel{\mathcal{N}}{=} 0\,, \label{k is affine only on N}
\end{align}
so $k^{a}$ is the tangent vector to the affinely parametrized null geodesic generators that generate the null hypersurface $\mathcal{N}$.

Another basis vector field is defined by
\begin{align}
    \ell^a \coloneqq (\partial _u)^a\,,
    \label{affine null l}
\end{align}
and it satisfies $\ell^a \gamma_{ab}=0 \,,\, \ell^a \beta_a =0$.
Then, the dual covector is
\begin{align}
\ell_a = g_{ab}\ell^b = -(dv)_a \,,
\end{align}
so $\ell_{a}$ is the normal vector of $v$-constant hypersurfaces $\set{\overline{\mathcal{N}}_{v}}$.
This is also the tangent vector to the affinely parametrized null geodesic generators:
\begin{align}
    &\ell^a \ell_a =0 \,,\\
    &\ell^b\nabla_b \ell_a = \ell^b\left(\partial_b \ell_a -\Gamma^c_{ba} \ell_c\right) 
    =\Gamma^v_{ua}=0\,.
\end{align}
$\ell^{a}$ is tangent to affinely parametrized null geodesics.
Here, $\ell^{a}$ is future directed because it satisfies
\begin{align}
    k^a \ell_a =-1\,.
\end{align}

The remaining basis vector is defined by 
\begin{align}
    m_A^a \coloneqq (\partial_A)^a\,,
\end{align}
and the vector fields $\set{k^{a}, \ell^{a}, m_{A}^{a}}$ form a complete basis. The covector is given by
\begin{align}
    m^A_a \coloneqq \gamma^{AB}m_B^b g_{ab}= \gamma^{AB}\left[-u\beta_B (dv)_a + \gamma_{BC}(d\varphi^C)_a\right]
    =(d\varphi^A)_a-u\beta^A(dv)_a \stackrel{\mathcal{N}}{=} (d\varphi^A)_a \,.
\end{align}

Let us introduce the cross-section $\mathcal{C}_{v}$, defined as the intersection of $\mathcal{N}$ with a $v$-constant surface $\overline{\mathcal{N}}_v$, that is, $\mathcal{C}_v \coloneqq \mathcal{N}\cap \overline{\mathcal{N}}_v $.
In this paper, we assume that $\mathcal{C}_{v}$ is compact and boundaryless.
Since $k^{a}$ and $\ell^a$ are normal null vector fields to $\mathcal{C}_{v}$, the induced metric on $\mathcal{C}_{v}$ can be defined by $g_{ab} + 2 k_{(a}\ell_{b)}$. Using the expression of the metric in the affinely parametrized Gaussian null coordinates, we find that the induced metric coincides with $\gamma_{ab}$,  
\begin{align}
g_{ab} + 2 k_{(a}\ell_{b)} = \gamma_{ab}\,.
\end{align}
Furthermore, based on this construction of the induced metric, the volume $(\mathrm{D}-2)$-form on $\mathcal{C}_{v}$ can be written as follows.
Let us write $\mathrm{det}(g_{ab}) \eqqcolon g\,,\mathrm{det}(\gamma_{ab}) \eqqcolon \gamma$, then $-g= \gamma$.
Hence the volume $\mathrm{D}$-form $\boldsymbol{\varepsilon}$ can be expressed as 
\begin{align}\notag
    \boldsymbol{\varepsilon}&= \sqrt{\gamma} du\wedge dv \wedge d\varphi^1\wedge d\varphi^2 \wedge \cdots \wedge d\varphi^{\mathrm{D}-2}
    \\ & \stackrel{\mathcal{N}}{=} \sqrt{\gamma} k \wedge \ell \wedge d\varphi^1 \wedge \cdots \wedge d\varphi ^{\mathrm{D}-2}\,.
\end{align}
Then, we define the volume $(\mathrm{D}-1)$-form $\bm{\varepsilon}^\mathcal{N}$ on $\mathcal{N}$ by
\begin{align}
\bm{\varepsilon} \stackrel{\mathcal{N}}{\equiv} -k \wedge \bm{\varepsilon}^\mathcal{N}\,,
\end{align}
and the volume $(\mathrm{D} - 2)$-form $\bm{\varepsilon}^{\mathcal{C}_v}$ on $\mathcal{C}_{v}$ by
\begin{align}
   \bm{\varepsilon}^\mathcal{N} \stackrel{\mathcal{C}_v}{\equiv}   -\ell \wedge \bm{\varepsilon}^{\mathcal{C}_v}\,.
\end{align}
In the abstract index notation, it is represented as
\begin{align}
    (\bm{\varepsilon})_{a_1\cdots a_\mathrm{D}} &\stackrel{\mathcal{N}}{=}-\mathrm{D} k_{[a_1}(\bm{\varepsilon}^{\mathcal{N}})_{a_2\cdots a_\mathrm{D}]}
    \stackrel{\mathcal{C}_v}{=} \mathrm{D} (\mathrm{D}-1)k_{[a_1}\ell_{a_2}(\bm{\varepsilon}^{\mathcal{C}_v})_{a_3\cdots a_\mathrm{D}]}\, .
\end{align}
We introduce the interior product $\iota_\chi$, which maps a $p$-form $\bm{\omega}$, for a fixed vector field $\chi^a$, to a $(p-1)$-form $\iota_{\chi} \bm{\omega}$ defined as $(\iota_\chi \bm{\omega})_{a_2\cdots a_{p}} = \chi^{a_1} (\bm{\omega})_{a_1 \cdots a_p} $. Thus, the volume forms $\bm{\varepsilon}^{\mathcal{N}}$ and $\bm{\varepsilon}^{\mathcal{C}_v}$ take the form
\begin{align}
 \bm{\varepsilon}^{\mathcal{N}} &\stackrel{\mathcal{N}}{=}  \iota_{\ell} \bm{\varepsilon}\,, \\
 \bm{\varepsilon}^{\mathcal{C}_v} &\stackrel{\mathcal{C}_v}{=}  \iota_{k} \bm{\varepsilon}^{\mathcal{N}} \stackrel{\mathcal{N}}{=}   \iota_{k} \iota_{\ell} \bm{\varepsilon}\,, 
\end{align}
or in the abstract index notation, 
\begin{align}
    &(\bm{\varepsilon}^{\mathcal{N}})_{a_2\cdots a_\mathrm{D}} \stackrel{\mathcal{N}}{=} \ell^{a_1}(\bm{\varepsilon})_{a_1\cdots a_\mathrm{D}}
    =(\bm{\varepsilon})_{ua_2\cdots a_\mathrm{D}}\,,
    \\ &(\bm{\varepsilon}^{\mathcal{C}_v})_{a_3\cdots a_\mathrm{D}} \stackrel{\mathcal{C}_v}{=}
    k^{a_2} (\bm{\varepsilon}^\mathcal{N})_{a_2\cdots a_\mathrm{D}}
    \stackrel{\mathcal{N}}{=}  \ell^{a_1}k^{a_2}(\bm{\varepsilon})_{a_1\cdots a_\mathrm{D}}
    =(\bm{\varepsilon})_{uva_3\cdots a_\mathrm{D}}\,.
\end{align}

\subsection{Stationary Black Hole Backgrounds and the Boost Weight Analysis\label{subsec: bh}}
In this section, we focus on a stationary black hole background with a Killing vector $\xi^{a}$, which has a bifurcate Killing horizon, i.e., there are two intersecting Killing horizons $\mathcal{H}^{+}$ and $\mathcal{H}^{-}$ associated with the Killing vector $\xi^{a}$.
The intersection of the two Killing horizons, $\mathcal{B}\coloneqq \mathcal{H}^+ \cap \mathcal{H}^- \neq \varnothing$, is called the bifurcation surface.  
The Killing vector $\xi^{a}$ vanishes on the bifurcation surface $\mathcal{B}$, $\xi^{a} \stackrel{\mathcal{B}}{=} 0$.
We assume that all background fields respect the Killing symmetry as well as the metric.
Denoting the metric $g_{ab}$ and other matter fields $\psi$ collectively as $\phi$, this condition is expressed as
\begin{align}
\mathsterling_\xi \phi = 0\,.
\label{Killing symmetry}
\end{align}

Let us introduce the affinely parametrized Gaussian null coordinates with identifying $\mathcal{N}$ as a Killing horizon $\mathcal{H}^{+}$, thus,
\begin{align}
  \mathcal{N} \equiv \mathcal{H}^+  \,.
    \label{future horizon}
\end{align}
We also choose the origin of the $v$-coordinate so that $v = 0$ corresponds to the past Killing horizon $\mathcal{H}^{-}$:
\begin{align}
    \overline{\mathcal{N}}_0 \equiv \mathcal{H}^-\,.
    \label{past horizon}
\end{align}
Then, introducing the surface gravity $\kappa$ by
\begin{align}
 (\nabla_{[a} \xi_{b]})(\nabla^a \xi^b) \stackrel{\mathcal{N}}{=} -2\kappa^2\,,
\label{surface gravity}
\end{align}
and the Killing vector $\xi^{a}$ can be expressed in the affinely parametrized Gaussian null coordinates as
\begin{align}
    \xi^a = \kappa\left(v k^a -u\ell^a \right)\,,
    \label{Killing in AGNCs}
\end{align}
and implies that the functions $\alpha,\, \beta_{A}$, and $\gamma_{AB}$ depend on the coordinates $u$ and $v$ through the combination $u v$.
From this expression \eqref{Killing in AGNCs}, one can directly confirm that the following properties hold:
\begin{align}
    & \xi_a \xi^a \stackrel{\mathcal{H}^+}{=} 0\,,
    \label{xi is null}\\
    &\xi^b \nabla_b \xi^a   \stackrel{\mathcal{H}^+}{=} \kappa \xi^a \,,
    \label{xi is null generator}\\
    &\xi^a \stackrel{\mathcal{B}}{=}0\,.
    \label{xi vanish on B}
\end{align}

Here, we review the concept of boost weight\footnote{
Our definition of the boost weight has the opposite sign comparing with Refs.~\cite{Wall:2015raa,Wall:2024lbd,Hollands:2022fkn,Biswas:2022grc}.  
}
based on Refs.~\cite{Wall:2015raa,Wall:2024lbd,Hollands:2022fkn,Biswas:2022grc}
associated with the affinely parametrized Gaussian null coordinates, which is also related to the Killing vector introduced above. 
Let us consider a $(n, n^\prime)$ tensor field $T^{a_1\cdots a_n}{}_{b_1\cdots b_{n^\prime}}$. Then, by introducing the null local Lorentz basis by $\set{(e_{\mu})^{a}} \coloneqq \set{(\partial_{\mu})^{a}} = \set{k^{a}, \ell^{a}, m_{A}^{a}}$ and their dual basis $\set{(e^{\mu})_{a}} \coloneqq \set{(dx^{\mu})_{a}}$, the components of this tensor in this local Lorentz frame can be expressed as
\begin{align}
    T^{\mu_1 \cdots \mu_n}{}_{\nu_1\cdots \nu_{n^\prime}}
    = T^{a_1\cdots a_n}{}_{b_1\cdots b_{n^\prime}}(e^{\mu_1})_{a_1} \cdots (e^{\mu_n})_{a_n} (e_{\nu_1})^{b_1}\cdots (e_{\nu_{n^\prime}})^{b_{n^\prime}}\,.
\end{align}
We now introduce another set of local Lorentz bases, obtained by performing a boost on the original basis, as follows:
\begin{align}
    \begin{cases}
        (e^\prime_u)^a \coloneqq \mathrm{e}^{\sigma}(e_u)^a\\
        (e^\prime_v)^a \coloneqq \mathrm{e}^{-\sigma}(e_v)^a \\
        (e^\prime_{A})^{a} \coloneqq (e_{A})^{a}
    \end{cases} \,,\, 
    \begin{cases}
        (e^{\prime u})_a \coloneqq \mathrm{e}^{-\sigma}(e^u)_a\\
        (e^{\prime v})_a \coloneqq \mathrm{e}^{\sigma}(e^v)_a\\
        (e^{\prime A})_a \coloneqq (e^A)_a
    \end{cases}
,
\end{align}
where $\sigma$ is a constant.
Then, the components associated with the new basis can be written as  
\begin{align}
    (T^\prime)^{\mu_1 \cdots \mu_n}{}_{\nu_1\cdots \nu_{n^\prime}}
    = T^{a_1\cdots a_n}{}_{b_1\cdots b_{n^\prime}}(e^{\prime \mu_1})_{a_1} \cdots (e^{\prime \mu_n})_{a_n} (e^{\prime}_{\nu_1})^{b_1}\cdots (e^{\prime}_{\nu_{n^\prime}})^{b_{n^\prime}}\, .
\end{align}
Subsequently, the boost weight $w$ of each component in the local Lorentz frame $T^{\mu_1 \cdots \mu_n}{}_{\nu_1\cdots \nu_{n^\prime}}$ can be defined as
\begin{align}
   (T^\prime)^{\mu_1 \cdots \mu_n}{}_{\nu_1\cdots \nu_{n^\prime}}
    =\mathrm{e}^{w\sigma}T^{\mu_1 \cdots \mu_n}{}_{\nu_1\cdots \nu_{n^\prime}}\,.
\end{align}
By the definition, $w$ can be obtained by counting the number of $u$ and $v$ indices, 
\begin{align}
    w= \# {}^{v}-\# {}^{u}+\# {}_{u} -\# {}_{v},
    \label{boost weight for tensor field}
\end{align}
where $\#^P$ and $\#_P$ denote the number of upper and lower indices labeled by $P$, respectively.

Let us consider a diffeomorphism corresponding to the scaling transformation $(u, v, \varphi^A) \mapsto (\mathrm{e}^{- \sigma} u, \mathrm{e}^{\sigma} v, \varphi^A)$.
The generator associated with this diffeomorphism can be represented using the Killing vector $\xi^a$.
The Lie derivative of the tensor associated with this diffeomorphism can be evaluated as 
\begin{align}
    \mathsterling_{\xi} T^{a_1\cdots a_n}{}_{b_1\cdots b_{n^\prime}} &= \xi^c \partial_c T^{a_1\cdots a_n}{}_{b_1\cdots b_{n^\prime}} \notag\\
&\qquad     -\sum_{i=1}^n T^{a_1 \cdots c \cdots a_n}{}_{b_1\cdots b_{n^\prime}} \partial_c \xi^{a_i} + \sum_{j=1}^{n^\prime} T^{a_1\cdots a_n}{}_{b_1 \cdots c \cdots b_{n^\prime}} \partial_{b_j}\xi^c \,.
\label{Lie derived T}
\end{align}
In the Lorentz frame, the components can be expressed as 
\begin{align}
(\mathsterling_{\xi} T)^{\mu_1\cdots \mu_n}{}_{\nu_1\cdots \nu_{n^\prime}}&= \xi^\mu \partial_\mu T^{\mu_1\cdots \mu_n}{}_{\nu_1\cdots \nu_{n^\prime}} \notag\\
&\qquad     -\sum_{i=1}^n \kappa T^{\mu_1 \cdots \alpha \cdots \mu_n}{}_{\nu_1\cdots \nu_{n^\prime}} \left(\delta^v{}_\alpha \delta_v{}^{\mu_i} - \delta^u{}_\alpha \delta_u{}^{\mu_i}\right) \notag\\
&\qquad + \sum_{j=1}^{n^\prime}\kappa T^{\mu_1\cdots \mu_n}{}_{\nu_1 \cdots \alpha \cdots \nu_{n^\prime}} \left(\delta^v{}_{\nu_j}\delta_v{}^{\alpha} - \delta^u{}_{\nu_j} \delta_u{}^{\alpha}\right)
\notag \\ &= \xi^\mu \partial_\mu T^{\mu_1\cdots \mu_n}{}_{\nu_1\cdots \nu_{n^\prime}} -\kappa w T^{\mu_1\cdots \mu_n}{}_{\nu_1\cdots \nu_{n^\prime}}\,.
\end{align} 
If the tensor $T^{a_1\cdots a_n}{}_{b_1\cdots b_{n^\prime}}$ is stationary, the Lie derivative associated with $\xi^{a}$ vanish. 
Hence, we obtain,
\begin{align}
 \xi^\mu \partial_\mu T^{\mu_1\cdots \mu_n}{}_{\nu_1\cdots \nu_{n^\prime}} = \kappa w T^{\mu_1\cdots \mu_n}{}_{\nu_1\cdots \nu_{n^\prime}}.
\end{align}
By solving this differential equation on the $u = 0$ hypersurface $\mathcal{H}^{+}$,
we obtain, for any stationary tensor field,
\begin{align}
    T^{\mu_1\cdots \mu_n}{}_{\nu_1\cdots \nu_{n^\prime}} (v,0,\varphi^A) = C(\varphi^A)v^w
\end{align}
where $C(\varphi^A)$ is an arbitrary function.
Thus, the $v$-dependence on the hypersurface $\mathcal{H}^{+}$ can be specified by its boost weight $w$. In addition, if $T^{a_1a_2\cdots a_n}{}_{b_1b_2\cdots b_{n^\prime}}$ is a regular tensor on $\mathcal{B}$, $C(\varphi^A)=0$ is required for $w < 0$.


\section{\label{sec: review on HWZ}A Review of the Dynamical Black Hole Entropy}
The Hollands--Wald--Zhang (HWZ) entropy is defined on an arbitrary cross-section $\mc{C}_v$ of the background Killing horizon $\mc{H^+}$, and is valid to leading order for non-stationary perturbations of a stationary black hole background with the bifurcation surface $\mathcal{B}$, using the Noether charge approach in the covariant phase space formalism \cite{Hollands:2024vbe}.
The entropy has a non-trivial correction term from non-stationary perturbations.
In particular, Ref.~\cite{Hollands:2024vbe} also show that, in the case of the Einstein--Hilbert action, the HWZ entropy does not coincide with
the Bekenstein--Hawking entropy, which is proportional to the area of the event horizon, nor the areal entropy of the background Killing horizon, but it is proportional to the area of the apparent horizon. 
An alternative proof of this relation was given by Visser and Yan~\cite{Visser:2024pwz}, which we essentially follow in this work.

The HWZ entropy admits a simple geometric expression in general relativity: the areal entropy evaluated on the apparent horizon. However, it remains unclear whether a similarly simple geometric characterization exists in general diffeomorphism invariant theories of gravity, apart from a few special cases \cite{Kong:2024sqc}.
In this paper, we propose a general framework in which the HWZ entropy on a horizon cross-section can be characterized as the Wall entropy evaluated on a distinguished surface relevant to the dynamical horizon structure, which we identify as the entropic marginally outer trapped surface (E-MOTS). 
The precise definition and construction of the E-MOTS will be given in the next section.
To prepare for this discussion, we begin by briefly reviewing the HWZ entropy.

\subsection{\label{subsec: review on cov phase space}The Covariant Phase Space Formalism}
In this subsection, we begin by introducing the Noether charge method based on the covariant phase space formalism, following Ref.~\cite{Iyer:1994ys}, and present the fundamental identity derived from it.
Let us consider a diffeomorphism invariant theory on $(\mc{M},g_{ab})$ with a Lagrangian $\mr{D}$-form $\bm{L}$, which can generally be written by
\begin{align}
    \bm{L} = \bm{\epv} L(g_{ab},R_{abcd},\nabla_{a_1}R_{abcd},\nabla_{(a_1}\nabla_{a_2)}R_{abcd},\cdots, \nabla_{(a_1}\cdots \nabla_{a_m)}R_{abcd} ,\psi,\nabla_{a_1} \psi ,\dots ,\nabla_{(a_1}\cdots \nabla_{a_\ell)} \psi )\,.
    \label{general covariant L}
\end{align}
Here, $(\mc{M}, \bm{\epv})$ is the volume $\mr{D}$-form associated with $g_{ab}$ and $\nabla$ and $R_{abcd}$ denote the covariant derivative and the Riemann curvature tensor with respect to the metric $g_{ab}$ respectively.
$\psi$ represents any matter fields in the theory.

Let $\phi$ denote the collection of dynamical fields $g_{ab}$ and $\psi$. Then, the first-order variation of $\bm{L}$ can be written as
\begin{align}
    \delta \bm{L}(\phi)= \bm{E}(\phi)\delta \phi + d \bm{\theta}(\phi,\delta\phi)\,.
\label{1st order variation of L}
\end{align}
Note that $\bm{E}(\phi) \delta \phi$ is understood as the sum of the variation with respect to all the dynamical variables, thus,
\begin{align}
\bm{E}(\phi)\delta \phi = \bm{\varepsilon} \left( E_{g}^{ab} \delta g_{ab} + E_{\psi} \delta \psi  \right).
\end{align}
The first term on the right-hand side of Eq.~\eqref{1st order variation of L} gives the equations of motion (i.e., the on-shell conditions) as $\bm{E}(\phi)=0$, and the second term is called the boundary term, which is expressed as the exterior derivative of the symplectic potential $(\mr{D}-1)$-form $\bm{\theta}$.
The symplectic potential serves as a fundamental building block in defining charges in the covariant phase space formalism.

We define the symplectic current $(\mr{D}-1)$-form $\bm{\omega}$ with two independent field variations $\delta_1 \phi$ and $\delta_2 \phi$ as
\begin{align}
    \bm{\omega} (\phi,\delta_1 \phi, \delta_2 \phi) \coloneqq \delta_1 \bm{\theta}(\phi,\delta_2 \phi)- \delta_2 \bm{\theta}(\phi,\delta_1 \phi )\,.
\label{presymplectic current}
\end{align}
If the perturbations satisfy the both ``linearized" equations of motion $\delta_1 \bm{E} =0$ and $\delta_2 \bm{E} =0$, the symplectic current is conserved, i.e., 
\begin{align}
    d \bm{\omega} =0\,.
    \label{domega equal zero}
\end{align}
This property is essential to define the symplectic 2-form on the covariant phase space:
\begin{align}
    \Omega_\Sigma (\phi,\delta_1 \phi,\delta_2 \phi) \coloneqq \int_\Sigma \bm{\omega}(\phi,\delta_1 \phi,\delta_2 \phi)\,.
    \label{gen symplectic form}
\end{align}
where $\Sigma$ is a Cauchy surface.

Let $\chi^a$ be an arbitrary smooth vector field on $\mc{M}$, and consider a special variation $\delta_\chi \phi \equiv \lie{\chi} \phi$ induced by a diffeomorphism generated by $\chi^a$.
We define the Noether current $(\mr{D}-1)$-form associated with the diffeomorphism generated by $\chi^a$ as 
\begin{align}
    \bm{J}_\chi \coloneqq \bm{\theta}(\phi,\lie{\chi} \phi) -\iota_\chi \bm{L}(\phi)\,.
    \label{def Noether current}
\end{align}
Using the fact that $\delta_\chi \bm{L}= d\iota_\chi \bm{L} $ and also $\delta_\chi \bm{L} = \bm{E}\lie{\chi} \phi + d \bm{\theta}(\phi,\lie{\chi} \phi )$, we obtain
\begin{align}
    d\bm{J}_\chi = -\bm{E}(\phi) \lie{\chi} \phi \,,
    \label{d Noether current}
\end{align}
so $\bm{J}_\chi$ is closed when the equations of motion hold.
From this observation, we can express $\bm{J}_\chi$ as
\begin{align}
    \bm{J}_\chi= d \bm{Q}_\chi + \bm{C}_\chi\,.
    \label{def Q and C}
\end{align}
Here, $\bm{C}_\chi$ is linear in $\chi^a$ and vanishes under the equations of motion.
$\bm{Q}_\chi$ is the Noether charge $(\mr{D}-2)$-form associated with the diffeomorphism generated by $\chi^a$.
The detailed expressions are given in App.~\ref{app: Noether Charges}. 

From the definition of the Noether current \eqref{def Noether current} and its Lie derivative 
\begin{align}
    \mc{L}_\chi \bm{\theta}(\phi,\delta \phi)= \iota_\chi d\bm{\theta}(\phi,\delta \phi) + d \left(\iota_\chi \bm{\theta}(\phi,\delta \phi)\right)\,,
\end{align}
we obtain
\begin{align}
    \delta \bm{J}_\chi& = \delta \bm{\theta}(\phi,\mc{L}_\chi \phi) -\iota_\chi \left(\bm{E}(\phi) \delta \phi + d\bm{\theta}(\phi,\delta \phi)\right)\nn
    &=\delta \bm{\theta}(\phi,\mc{L}_\chi \phi) -\mc{L}_\chi \bm{\theta}(\phi,\delta \phi) + d\iota_\chi\bm{\theta}(\phi,\delta \phi)-\iota_\chi \bm{E}(\phi) \delta \phi\,. \label{delta J another form}
\end{align}
Hence, we can rewrite Eq.~\eqref{delta J another form} using 
Eqs.~\eqref{presymplectic current} 
and~\eqref{def Q and C}
as
\begin{align}
    \bm{\omega} (\phi,\delta \phi,\lie{\chi}\phi)  &= \delta \bm{J}_\chi(\phi,\delta\phi) -d \iota_\chi \bm{\theta}(\phi,\delta\phi)+\iota_\chi\bm{E}(\phi)\delta\phi 
    \nn &= d \left[ \delta \bm{Q}_\chi(\phi,\delta \phi) -\iota_\chi \bm{\theta}(\phi,\delta\phi) \right] 
    +\iota_\chi\bm{E}(\phi)\delta\phi + \delta \bm{C}_\chi(\phi)\,.
    \label{fundamental id}
\end{align} Eq.~\eqref{fundamental id} is known as the fundamental identity of the covariant phase space formalism.

\subsection{\label{subsec: HWZ}The Hollands--Wald--Zhang Entropy}
In this subsection, using the results obtained so far, we define the HWZ entropy and explain its structure. Then, we explicitly present its relation to the Iyer--Wald entropy \cite{Iyer:1994ys} and the Wall entropy \cite{Wall:2015raa}.

\subsubsection{\label{subsub: geometrical setup}Our Setup}

In the following, we focus on a black hole spacetime that is approximately stationary in the sense that the physical metric $g_{ab}$ of interest can be expressed as a perturbation around a stationary black hole metric $\bg{g}_{ab}$.
Throughout the following, we indicate background quantities by denoting them with a small circle. We focus on the case where the background metric is a solution to the equations of motion which are derived from the Lagrangian, thus, $ \bg{\bm{E}} = 0$ and it also implies $\bg{\bm{C}}_{\chi} = 0$.

To formulate the perturbative analysis rigorously, 
we introduce a one-parameter family of metrics $g_{ab}(\epsilon)$ labeled by $\epsilon$, which we denote by the full metric in this paper, 
that reduces to the stationary background metric $\stackrel{\circ}{g}_{ab}$ at $\epsilon = 0$ and the physical metric $g_{ab}$ of interest at $\epsilon = 1$, thus
\begin{align}
g_{ab}(0) = \bg{g}_{ab}\,, \qquad g_{ab}(1) = g_{ab}\,. 
\end{align}
In this paper, we focus on the first-order perturbations in $\epsilon$. Therefore, the metric $g_{ab}(\epsilon)$ can be expanded as 
\begin{align}
    g_{ab}(\epsilon) = \stackrel{\circ}{g}_{ab}+ \epsilon \delta g_{ab} + \mathcal{O}(\epsilon^2)\,.
\end{align}
We note that, although we systematically perform the perturbative calculations as a Taylor expansion in $\epsilon$, the validity of the approximation after setting $\epsilon = 1$ relies on the smallness of $\delta g_{ab}$ itself.

There are gauge degrees of freedom associated with the choice of a one-parameter family of metrics as well as the identification of the background spacetime $\bg{g}_{ab}$; we can use another background metric $\bg{\hat{g}}_{ab}$ which is related to a perturbative diffeomorphism from $\bg{g}_{ab}$, namely, $\bg{\hat{g}}_{ab} = ~\bg{g}_{ab} + \mathsterling_{\eta} \bg{g}_{ab} + \cdots$, where $\eta^a$ is the generator of a diffeomorphism between the two background metrics.
 As a gauge choice, we impose the following condition:
Let $\mathcal{N}$ be a null hypersurface which can be identified with a Killing horizon associated with some background stationary black hole, $\mc{N} \equiv \mc{H}^{+}$. We refer to such $\mathcal{N}$ as an \textit{background Killing horizon}. 
Note that a background Killing horizon is not unique, since any null hypersurface that differs from it by a perturbatively small amount is also regarded as a background Killing horizon. We require that the deformation by $\epsilon$ preserves the metric ansatz in the affinely parametrized Gaussian null coordinates associated with a given background Killing horizon $\mathcal{N}$. Thus, the metric $g_{ab}(\epsilon)$ can be expressed as 
\begin{align}
g_{ab}(\epsilon) = - 2 (dv)_{(a}  \left[(du)_{b)} +\frac{u^2}{2} \alpha(\epsilon) (dv) _{b)} + u \beta_{b)}(\epsilon) \right]+ \gamma_{ab}(\epsilon)\,.  
    \label{one para rep for AGNCs}
\end{align}
Unless otherwise specified, the indices of tensors are raised and lowered using the metric $g_{ab}(\epsilon)$ and its inverse $g^{ab}(\epsilon)$.
For example, although $k^{a} = (\partial_{v})^{a}$ is independent of $\epsilon$ by definition, $k_{a} = g_{ab}(\epsilon) k^{b} = - (du)_{a} - u^2 \alpha(\epsilon) (d v)_{a} - u \beta_{a}(\epsilon)$ depends on $\epsilon$.

If matter fields $\psi$ are present, we also impose appropriate gauge conditions on their perturbations.
Since our main focus is on the geometric characterization of the HWZ entropy, we shall, for the remainder of this paper, restrict our attention to the case where the metric is the only dynamical field in the Lagrangian.
Thus, $\phi = g_{ab}$ and, for example, $\bm{E}(\phi) \delta \phi = \bm{\varepsilon}E_{g}^{ab} \delta g_{ab}.$
The following discussion is based on that of Visser and Yan \cite{Visser:2024pwz}, where general bosonic matter fields are included in the Lagrangian. Accordingly, our analysis can be straightforwardly extended to such cases\footnote{The covariant phase space formalism including spinor fields has been discussed in Ref.~\cite{Prabhu:2015vua}. Extending our analysis to such a setting would be an interesting direction for future works.}.

We define the vector field $\xi^{a}$ by \eqref{Killing in AGNCs} with $\kappa$ denoting the surface gravity of the background spacetime\footnote{
We can also consider the perturbation of the vector field $\xi^a$, however, as in the analysis given in Ref.~\cite{Visser:2024pwz}, it does not affect the leading order analysis of the HWZ entropy.
}. 
Since we do not impose the stationary symmetry to the perturbations, $\xi^{a}$ is not a Killing vector of the full metric $g_{ab}(\epsilon)$.
In particular, the present discussion includes cases where 
\begin{align}
    \delta (\mathsterling_\xi g(\epsilon))\neq 0\,,
    \label{def of non-stationary perturbation}
\end{align}
and we are especially interested in such a non-stationary perturbation.

An important consequence of the boost weight argument is that a quantity with negative boost weight must be at least first-order in $\epsilon$.
Similarly, a quantity that can be expressed as a product of two quantities both of which have negative boost weight is at least quadratic order in $\epsilon$ and can be neglected in the first-order analysis.

Throughout the rest of this section, we consider the full metric $g_{ab}(\epsilon)$. For notational simplicity, we will suppress the $\epsilon$-dependence whenever it does not lead to confusion.

\subsubsection{\label{subsub: structure of the symplectic potential}The Structure of Symplectic Potentials}

Integrating the fundamental identity Eq.~\eqref{fundamental id} with $\chi^a \equiv \xi^a$ over a portion of the horizon $\Delta \mc{H}^+$, and imposing the background on-shell conditions, we obtain
\begin{align}
0=\int_{\Delta \mc{H}^+} \bm{\omega}(\bg{g},\delta g, \mathsterling_\xi \!\!\bg{g}  ) &= \int_{\Delta \mc{H}^+}d \left[\delta \bm{Q}_\xi -\iota_\xi \bm{\theta} \right] + \int_{\Delta \mc{H}^+} \delta \bm{C}_\xi
    \nn &= \int_{\partial(\Delta \mc{H}^+)}\left[\delta \bm{Q}_\xi -\iota_\xi \bm{\theta} \right]
    + \int_{\Delta \mc{H}^+} \delta \bm{C}_\xi\,.
\label{integrate fundamental id on H}
\end{align}
In the first equality, we have used the fact that $\bm{\omega}(\bg{g},\delta g, \mathsterling_\xi \!\!\bg{g} )$ is linear in $\mathsterling_\xi \!\!\bg{g} $ and vanishes due to $\mathsterling_\xi \!\!\bg{g}  =0$.
If the pullback of the symplectic potential to $\mc{H}^+$, denoted by $\underline{\bm{\theta}}$, can be written as a total variation, namely
\begin{align}
    \underline{\bm{\theta}}(\bg{g},\delta g) = \delta \bm{B}_{\mc{H}^+}\,,
    \label{B theta relation to prove}
\end{align}
the HWZ entropy $(\mr{D}-2)$-form can be defined by
\begin{align}
\bm{s}^\mr{HWZ} \coloneqq \frac{2\pi}{\kappa \hbar} \left[\underline{\bm{Q}_\xi} -\iota_\xi \bm{B}_{\mc{H}^+}\right]  \,.
\label{def HWZ entropy form}  
\end{align}
Here, $(\mathrm{D}-2)$-forms with the underline denote that they are pulled back to an arbitrary cross-section $\mc{C}_v$ of the horizon $\mc{H}^+$.
The HWZ entropy on $\mc{C}_v$ is defined as 
\begin{align}
    S_\mr{HWZ} (\mc{C}_v) \coloneqq \int_{\mc{C}_v} \bm{s}^\mr{HWZ} = \frac{2\pi}{\kappa \hbar} \int_{\mc{C}_v}    \left[\underline{\bm{Q}_\xi} -\iota_\xi \bm{B}_{\mc{H}^+}\right] \,.
    \label{def HWZ entropy}
\end{align}
This definition is motivated by the definition of the localized charge (also known as the Wald-Zoupas charge) and its associated flux formula \cite{Wald:1999wa,Chandrasekaran:2018aop,Grant:2021sxk}.

In what follows, we show that the pullback of the symplectic potential is indeed the total variation of a quantity $\bm{B}_{\mc{H}^+}$, which is locally and covariantly constructed from the metric and the Riemann curvature tensor and/or its covariant derivatives.
As in Appendix \ref{app: Charges}, the symplectic potential up to Jacobson-Kang-Myers (JKM) ambiguities \cite{Jacobson:1993vj}\footnote{
As we will see later, these ambiguities do not affect our discussion.
} is given by
\begin{align}
    \bm{\theta}(\bg{g},\delta g)=\bg{\bm{\epv}}_a \biggl[2\!\bg{{E}}_R^{\,abcd}\nabla_d\delta g_{bc}+\bg{{S}}^{\,abc}\delta g_{bc}+\sum_{i=0}^{m-1}\bg{{T}}_{(i)}^{\,abcdeb_1\cdots b_i}\delta\bigl(\nabla_{(b_1}\cdots\nabla_{b_i)}R_{bcde}\bigr)
    \biggr]\,,
    \label{theta}
\end{align}
where we have used the following notation
\begin{align}
    (\bm{\epv}_a V^a)_{a_2\cdots a_\mr{D}} \equiv (\bm{\epv})_{aa_2\cdots a_\mr{D}} V^a= (\iota_V \bm{\epv})_{a_2\cdots a_\mr{D}}\,.
    \label{def epv_a }
\end{align}
Here, $E_R^{abcd},\, S^{abc}$ and $T_{(i)}^{abcdeb_1\cdots b_i}$ are tensor fields locally and covariantly constructed from metric and Riemann curvature and/or its covariant derivatives.
Their origin is briefly explained in the appendix \ref{app: Charges}.
Taking the pullback of the symplectic potential to the background Killing horizon $\mc{H}^+$, we obtain
\begin{align}
    \underline{\bm{\theta}}(\bg{g},\delta g)= \bg{\bm{\epv}}^{\mc{H}^+} &\biggl[2\!\bg{{E}}_R^{\,ubcd}\nabla_d\delta g_{bc}+\bg{{S}}^{\,ubc}\delta g_{bc}+\sum_{i=0}^{m-1}\bg{{T}}_{(i)}^{\,ubcdeb_1\cdots b_i}\delta\bigl(\nabla_{(b_1}\cdots\nabla_{b_i)}R_{bcde}\bigr)
    \biggr]\,,\label{theta general pullback}
\end{align}
where $(\mathrm{D}-1)$-forms with the underline denote that they are pulled-back to the horizon $\mc{H}^+$ and $\bm{\epv}^{\mc{H}^+}$ is the volume $(\mathrm{D} - 1)$-form on $\mc{H}^+$. 

We now analyze the structure of these terms using the boost weight argument introduced in Sec.~\ref{subsec: bh}.
Since $\bg{E}_{R}^{\,abcd}, \bg{S}^{\,abc}$ and $\bg{T}_{(i)}^{\,abcdeb_{1} \cdots b_{i}}$ are constructed from the stationary background quantities, their components with negative boost weight must vanish.
For the $E_R$ term, the boost weight argument, together with the assumption of the regularity of $\bg{E}_R$ gives
\begin{align}
    \bg{\bm{\epv}}^{\mc{H^+}}\!\!\bg{E}_R^{\,ubcd}\nabla_d\delta g_{bc}
    \stackrel{\mc{H}^+}{=}2\bg{\bm{\epv}}^{\mc{H^+}}\!\!\bg{{E}}_R^{\,uABv}\delta K_{AB}
    \stackrel{\mc{H}^+}{=}\delta\Bigl(2\bm{\epv}^{\mc{H^+}}E_R^{uABv} K_{AB}\Bigr)\,.
        \label{delta ER}
\end{align}
In the last equality, we have used the fact that $K_{AB}= \frac{1}{2} \del_v \gamma_{AB}$ has the boost weight $-1$, which leads $\bg{K}_{\!\!AB}\stackrel{\mc{H}^+}{=}0$.
For the $S$ term, the boost weight argument, together with the assumption of the regularity of $\bg{S}$, implies
\begin{align}
    \bg{S}^{\,ubc}\delta g_{bc} \stackrel{\mathcal{H}^+}{=} \bg{S}^{\,uAB}\delta g_{AB}=0\,.
     \label{weght S}
 \end{align}
The last equality follows from the fact that $S^{uAB}$ has boost weight $-1$, and thus must vanish on the background spacetime. 
For the $T$ term, we decompose it into a sum of terms with different boost weight $w$, yielding
\begin{align}
 \sum_{i=0}^{m-1}\bg{{T}}_{(i)}^{\,ubcdeb_1\cdots b_i}\delta\bigl(\nabla_{(b_1}\cdots\nabla_{b_i)}R_{bcde}\bigr)
& = 
\displaystyle\sum_w\bg{{T}}_{(w-1)}\delta(\nabla\cdots\nabla R)_{(-w)} \notag\\
  & =\displaystyle \sum_{w\ge 1}\bg{{T}}_{(w-1)}\delta(\nabla\cdots\nabla R)_{(-w)}
  \nn & =\delta \left[\displaystyle\sum_{w\ge 1}{T}_{(w-1)}(\nabla\cdots\nabla R)_{(-w)}\right]\,.
    \label{weight T}
\end{align}
The second equality follows from the fact that $\bg{T}_{(w-1)}$ with negative boost weight must vanish, and in the last line, we have used the following fact
\begin{align}
    A_{(w\ge 0 )} \delta B_{(w<0)} = \delta \left[   A_{(w\ge 0 )} B_{(w<0)}\right] -  \delta A_{(w\ge 0 )} B_{(w<0)}
    \stackrel{\mc{H}^+}{=} \delta \left[   A_{(w\ge 0 )} B_{(w<0)}\right]\,.
\end{align}
We note that following is also hold:
\begin{align}
    \bg{\bm{\epv}}^{\mc{H}^+}\delta \left[\displaystyle\sum_{w\ge 1}{T}_{(w-1)}(\nabla\cdots\nabla R)_{(-w)}\right]
    =\delta \left[\bm{\epv}^{\mc{H}^+}\displaystyle\sum_{w\ge 1}{T}_{(w-1)}(\nabla\cdots\nabla R)_{(-w)}\right].
\end{align}

From the above discussion, we can conclude that the pullback of the symplectic potential can be expressed as a total variation of the following quantity:
\begin{align}
    \underline{\bm{\theta}} (\bg{g},\delta g)= \delta \bm{B}_{\mc{H}^+} 
    \,,
    \label{B theta relation}
\end{align}
where
\begin{align}
    \bm{B}_{\mathcal{H}^+}=\bm{\epv}^{\mc{H}^+} \left[4{E}_R^{uABv}K_{AB}
    +\displaystyle\sum_{w\ge 1} {T}_{(w-1)}(\nabla\cdots\nabla R)_{(-w)}
    \right]+\mc{O}(\ep^2) \,.
    \label{B expression}
\end{align}
Note that the quantity in brackets in Eq.~\eqref{B expression} has boost weight $-1$, which implies that
\begin{align}
\bg{\bm{B}}_{\mc{H}^+}=0 \,.
\label{B vanishes on bg bh}
\end{align}

\subsubsection{The HWZ Entropy and the JKM Ambiguities}
We have seen that the pullback of the symplectic potential to the background Killing horizon $\mc{H}^+$ becomes a total variation of $\bm{B}_{\mc{H}^+}$.  
Accordingly, we define the HWZ entropy $(\mr{D}-2)$-form as
\begin{align}
\bm{s}^\mr{HWZ} \coloneqq \frac{2\pi}{\kappa \hbar} \left[\underline{\bm{Q}_\xi} -\iota_\xi \bm{B}_{\mc{H}^+}\right],
\label{def HWZ form again}
\end{align}
and the HWZ entropy on a horizon cross-section $\mc{C}_v$ as
\begin{align}
S_\mr{HWZ}(\mc{C}_v) \coloneqq \int_{\mc{C}_v} \bm{s}^\mr{HWZ} = \frac{2\pi}{\kappa \hbar} \int_{\mc{C}_v} \left[\underline{\bm{Q}_\xi} -\iota_\xi \bm{B}_{\mc{H}^+}\right].
\label{def HWZ entropy again}
\end{align}

We briefly recall the Jacobson--Kang--Myers (JKM) ambiguities in the covariant phase space formalism (see Refs.~\cite{Jacobson:1993vj,Visser:2024pwz} or Appendix~\ref{app: Charges}).
In the definition of the Noether charge $(\mr{D}-2)$-form, there are JKM ambiguities under the transformation:
\begin{align}
 \bm{Q}_\xi \to \bm{Q}_\xi + \iota_\xi \bm{\lambda}(g) + \bm{Y}(g, \lie{\xi} g) + d \bm{Z}_\xi(g)\,.
 \label{Q ambiguities}
\end{align}
The ambiguity $\bm{\lambda}$ arises from the degrees of freedom to shift the Lagrangian by an exact form, $\bm{L} \to \bm{L} + d\bm{\lambda}$, which does not affect the equations of motion.  
The ambiguity $\bm{Y}$ stems from the definition of the symplectic potential $\bm{\theta}$, which is only fixed up to the addition of an exact form.
The ambiguity $\bm{Z}_\xi$ reflects the fact that $\bm{Q}_\xi$ is defined only up to a closed form, due to the identity $\bm{J}_\xi = d\bm{Q}_\xi + \bm{C}_\xi$.

These ambiguities also lead to the following transformation of $\bm{\theta}$:
\begin{align}
\bm{\theta}(g,\delta g) \to \bm{\theta}(g,\delta g) + \delta \bm{\lambda}(g) + d\bm{Y}(g,\delta g)\,.\label{ambiguities for theta}
\end{align}
Consequently, they induce an ambiguity in the variation of the HWZ entropy form:
\begin{align}
\delta\bm{s}^\mr{HWZ} \to \delta\bm{s}^\mr{HWZ} + \frac{2\pi}{\kappa \hbar} d\left[\iota_\xi \bm{Y}(g,\delta g) +\delta \bm{Z}_\xi (g)\right].
\end{align}
Since $\mc{C}_v$ has no boundary, the integral of any exact form over it vanishes.  
Therefore, the HWZ entropy is unaffected by the JKM ambiguities.

\subsubsection{The Relation to the Iyer--Wald Entropy and the Wall Entropy\label{subsub: Wall}}

Let us clarify the relation between the HWZ entropy and other entropy formulae, the Iyer--Wald entropy \cite{Iyer:1994ys}\footnote{We note that the term ``Iyer--Wald entropy" in this paper refers to the extension of the Wald entropy to an arbitrary cross-section,
as explained in the footnote \ref{fn: IY}.
}
and the Wall entropy \cite{Wall:2015raa}. Let us express the Noether charge form $\bm{Q}_\xi$ more explicitly. Up to the JKM ambiguities, the Noether charge form is given by
\begin{align}
    \bm{Q}_\xi 
    &= \bm{\epv}_{ab} \left[W^{abc}(g)\xi_c -E_R^{abcd} \nabla_{[c}\xi_{d]} \right]\,.
   \label{Q up to JKM}
\end{align}
Here, we have used the notation 
\begin{align}
    (\bm{\epv}_{ab} V^{ab})_{a_3\cdots a_\mr{D}} \equiv (\bm{\epv})_{aba_3\cdots a_\mr{D}}V^{ab}\,,
    \label{def epv_ab}
\end{align}
and $W^{abc}$ is a tensor field locally constructed from the metric, the curvature tensors, and/or their covariant derivatives.
Taking the pullback of the Noether charge form to the horizon cross-section $\mc{C}_v$, we get
\begin{align}
    \underline{\bm{Q}_\xi} = -2\bm{\epv}^{\mc{C}_v}({E}_R^{uvcd}\nabla_{[c}\xi_{d]}-{W}^{uvc}\xi_{c})\,.
    \label{pullback Q to the cross-section}
\end{align}
Since non-vanishing coefficients of $\nabla_{[a} \xi_{b]}$ on $\mc{H}^+$ are given by
\begin{align} &\nabla_{[u}\xi_{v]}=k^{[b}l^{a]}\nabla_{a}\xi_{b}\stackrel{\mathcal{H}^+}{=}\kappa\,,
    \label{nabla xi on H 1}
\\ &   \nabla_{[u}\xi_{A]}\stackrel{\mathcal{H}^+}{=}\frac{1}{2}\xi_u\partial_u(u\beta_A)\stackrel{\mathcal{H}^+}{=}\frac{1}{2}\xi_u\beta_A\,,
    \label{nabla xi on H 2}
\end{align}
we obtain
\begin{align}
    \underline{\bm{Q}_\xi} = -2\bm{\epv}^{\mc{C}_v}(2\kappa{E}_R^{uvuv} - \kappa v {E}_R^{uvuA}\beta_A + \kappa v {W}^{uvu})\,.
    \label{pullback Q expression}
\end{align}
Here, we have used the equation $\xi_{u} \stackrel{\mathcal{H}^{+}}{=} - \kappa v$.
Note that only the first term in Eq.~\eqref{pullback Q expression} survives on the background, as it is the only one with non-negative boost weight.
Using the expression \eqref{pullback Q expression}, the HWZ entropy form can be expressed as 
\begin{align}
    \bm{s}^{\text{HWZ}} &= 
    - \frac{8 \pi}{\hbar}
    \bm{\epv}^{\mc{C}_v} {E}_R^{uvuv} \notag\\
&\qquad
    - v \frac{2 \pi}{\hbar}
    \bm{\epv}^{\mc{C}_v} \left(- 2 {E}_R^{uvuA}\beta_A + 2 {W}^{uvu}
    + 4{E}_R^{uABv}K_{AB}
    + \displaystyle\sum_{w\ge 1} {T}_{(w-1)}(\nabla\cdots\nabla R)_{(-w)} \right) \notag\\
&\qquad +\mathcal{O}(\epsilon^2)
    . \label{sHWZ}
\end{align}
Here, we have used the equation $\iota_\xi \bm{\epv}^{\mc{H}^+} \stackrel{\mc{C}_v}{=} -\kappa v \bm{\epv}^{\mc{C}_v} $ in the derivation.

We define the Iyer--Wald entropy form $\bm{s}^\mr{IW}$ as
\begin{align}
    \bm{s}^\mr{IW} \coloneqq -\frac{8\pi}{\hbar} \bm{\epv}^{\mc{C}_v} E_R^{uvuv}\,,
    \label{SIW form}
\end{align}
and the Iyer--Wald (IW) entropy on $\mc{C}_v$ as
\begin{align}
    S_\mr{IW}(\mc{C}_v) \coloneqq  \int_{\mc{C}_v} \bm{s}^\mr{IW} =  - \frac{8\pi}{\hbar} \int _{\mc{C}_v} E_R^{uvuv} \bm{\epv}^{\mc{C}_v}\,.
    \label{def IW entropy} 
\end{align}
Then, comparing with Eq.~\eqref{sHWZ},
we can express the HWZ entropy form as
\begin{align}
    &    \bm{s}^\mr{HWZ} = \bm{s}^\mr{IW} -v \bm{\varsigma} + \mathcal{O}(\epsilon^2) \,,
        \label{SHWZ form}
\end{align}
where we defined the non-stationary correction term $\bm{\varsigma}$ by
\begin{align}
    \bm{\varsigma}  =  \bm{\epv}^{\mc{C}_v} \biggl[ \frac{4\pi}{\hbar}\left(2E_R^{uABv} K_{AB} -E_R^{uvuA} \beta_A + W^{uvu}\right)
           + \frac{2\pi}{\hbar} \sum_{w\ge 1} 
                T_{(w-1)}\left(\nabla \cdots R\right)_{(-w)} \biggr]\,.
                \label{SHWZ form expression}
\end{align}
Note that $\bg{\bm{\varsigma}}=0$ from the boot weight argument, which implies that in stationary eras, the HWZ entropy coincides with the IW entropy.
We also note that $(v \bm{\varsigma})|_{v=0} =0$, which implies that on the bifurcation surface $\mc{B}$, located at $v = 0$, the HWZ entropy agrees with the Wald entropy—defined as the IW entropy evaluated on $\mc{B}$. In particular, 
\begin{align}
    S_\mr{Wald} \coloneqq S_\mr{IW} (\mc{B}) =  S_\mr{HWZ} (\mc{B}) = - \frac{8\pi}{\hbar} \int _{\mc{B}} E_R^{uvuv} \bm{\epv}^{\mc{B}},
    \label{def Wald entropy}
\end{align}
up to first-order in $\epsilon$.
If we denote the non-stationary correction term by
\begin{align}
    \Xi(\mc{C}_v) = \int_{\mc{C}_v} \bm{\varsigma}\,,
    \label{Xi}
\end{align}
then the HWZ entropy can be written as
\begin{align}
    & S_\mr{HWZ}(\mc{C}_v) =  S_\mr{IW}(\mc{C}_v) -v \Xi(\mc{C}_v)+ \mathcal{O}(\epsilon^2)\,.
    \label{SHWZ decomposition}
\end{align}

The Wall entropy \cite{Wall:2015raa} is defined by the following relation in the first-order analysis of perturbations around a stationary black hole background:
\begin{align}
     \partial_v^2\delta S_\mr{Wall}(\mc{C}_v)\equiv  \frac{4\pi}{\hbar}\int_{\mathcal{C}_v}\bg{\bm{\epv}}^{\,\mc{C}_v}(\delta E_{g})_{ab}k^ak^b 
     = \frac{2\pi}{\kappa\hbar v} \int_{\mc{C}_v} \iota_k \delta \bm{C}_\xi \,,
    \label{Wall def}
\end{align}
where the second equality follows from Eq.~\eqref{C expression without matter}.
From the fundamental identity Eq.~\eqref{fundamental id} under the background equations of motion, we have
\begin{align}
    \underline{\iota_k \delta \bm{C}_\xi} = -\frac{\kappa \hbar}{2\pi} \iota_k d \delta \bm{s}^\mr{HWZ}
    =  -\frac{\kappa \hbar}{2\pi} \left[\mathsterling_k \delta \bm{s}^\mr{HWZ} -d \iota_k \delta \bm{s}^\mr{HWZ}\right],
    \label{rel HWZ and external matter}
\end{align}
and substituting this expression into Eq.~\eqref{Wall def}, 
we obtain
\begin{align}
    \partial_v^2\delta S_\mr{Wall}(\mc{C}_v) = -\frac{1}{v} \del_v \delta S_\mr{HWZ}(\mc{C}_v)\,.
    \label{rel Wall and HWZ}
\end{align} 
This equation is equivalent to
\begin{align}
    \partial_v\delta S_\mr{HWZ}(\mc{C}_v)& = -v \partial_v^2\delta S_\mr{Wall}(\mc{C}_v)
\nn & = \partial_v\left[(1-v\partial_v)\delta S_\mr{Wall}(\mc{C}_v)\right]\,,
\end{align}
so, integrating over $v$ with the boundary condition 
\begin{align}
    S_\mr{Wall}(\mc{B}) = S_\mr{HWZ}(\mc{B}) \Big ( = S_{\mr{IW}}(\mc{B})  = S_\mr{Wald} \Big) \,,
\end{align}
gives
\begin{align}
     S_\mr{HWZ}(\mc{C}_v)=(1-v\partial_v)S_\mr{Wall}(\mc{C}_v) + \mathcal{O}(\epsilon^2)\,.
    \label{relation between HWZ and Wall}
\end{align}
This equation is the key relation for providing our proposal for the geometric characterization of the HWZ entropy.

Hollands, Wald, and Zhang \cite{Hollands:2024vbe} have defined the Wall entropy form $\bm{s}^\mr{Wall}$ as
\begin{align}
    \bm{s}^\mr{Wall} \stackrel{~\mc{C}_v}{\coloneqq} \bm{s}^\mr{HWZ} + \frac{4\pi}{\kappa \hbar} P_\xi(g,\mathsterling_\xi g) \bm{\epv}^{\mc{C}_v} + \mathcal{O}(\epsilon^2)\,,
    \label{def s Wall}
\end{align}
where $P_{\xi}$ is defined by
\begin{align}
    P_\xi(g,  \lie{\xi} g) \coloneqq -\sum_{i=0}^{k-2} \tilde{U}_{(i)}^{c_1 \cdots c_i d e} \nabla_{\left(c_1\right.} \cdots \nabla_{\left.c_i\right)} \lie{\xi} g_{d e} \,.
    \label{P explicit form}
\end{align}
Here, $\tilde{U}$ is defined through the relation 
\begin{align}
  \xi_a\xi_b\sum_{i=0}^k U_{(i)}^{a b c_1 \cdots c_i d e} \nabla_{\left(c_1\right.} \cdots \nabla_{\left.c_i\right)} \delta g_{d e}
    \stackrel{\mc{H}^+}{=}
    \sum_{i=0}^{k-2} \tilde{U}_{(i)}^{c_1 \cdots c_i d e} \nabla_{\left(c_1\right.} \cdots \nabla_{\left.c_i\right)} \delta g_{d e} 
\end{align}
with $U$ defined by
\begin{align}
    \delta E_g^{a b}\eqqcolon \sum_{i=0}^k U_{(i)}^{a b c_1 \cdots c_i d e} \nabla_{\left(c_1\right.} \cdots \nabla_{\left.c_i\right)} \delta g_{d e}\,.
\end{align}
Note that $\bg{P}_\xi = 0$.
The Wall entropy form defined by Eq.~\eqref{def s Wall} actually the entropy form for the Wall entropy, thus, it satisfies
\begin{align}
    S_\mr{Wall} (\mc{C}_v) = \int_{\mc{C}_v} \bm{s}^\mr{Wall}\,.
    \label{good property for s Wall}
\end{align}

By rewriting the HWZ entropy by the IW entropy through \eqref{SHWZ form}, the Wall entropy form can be expressed as 
\begin{align}
    \bm{s}^\mr{Wall} \stackrel{\mc{C}_v}{=} \bm{s}^\mr{IW} + \bm{s}^P - v \bm{\varsigma}  \,,
    \label{Wall form}
\end{align}
where we defined $\bm{s}^P$ on the cross-section of the horizon $\mc{C}_v$ as
\begin{align}
    \bm{s}^P \coloneqq  \frac{4\pi}{\kappa \hbar} P_\xi(g,\mathsterling_\xi g) \bm{\epv}^{\mc{C}_v}.
    \label{P entropy}
\end{align}
Therefore, the Wall entropy can be expressed as 
\begin{align}
    \bm{s}^{\text{Wall}} &= 
    - \frac{8 \pi}{\hbar}
    \bm{\epv}^{\mc{C}_v} {E}_R^{uvuv} + \frac{4\pi}{\kappa \hbar} P_\xi(g,\mathsterling_\xi g) \bm{\epv}^{\mc{C}_v} \notag\\
&\qquad
    - v \frac{2 \pi}{\hbar}
    \bm{\epv}^{\mc{C}_v} \left(- 2 {E}_R^{uvuA}\beta_A + 2 {W}^{uvu}
    + 4{E}_R^{uABv}K_{AB}
    + \displaystyle\sum_{w\ge 1} {T}_{(w-1)}(\nabla\cdots\nabla R)_{(-w)} \right) \notag\\
   & \qquad + \mathcal{O}(\epsilon^2)\,.
   \label{swall}
\end{align}
Note that, at least when the Lagrangian is a function of the Riemann tensor, the Wall entropy is known to be equivalent to the Dong--Camps entropy \cite{Dong:2013qoa, Camps:2013zua}. 

In a stationary regime, $\bm{s}^P$ and $\bm{\varsigma}$ vanish, and the HWZ entropy coincides with both the Wall and Iyer--Wald entropies.
Moreover, if one considers an auxiliary, fully stationary spacetime that includes a stationary portion of the dynamical spacetime of interest, these entropies coincide with those evaluated on the bifurcation surface of the auxiliary spacetime, that is, the Wald entropy.
Therefore, we obtain the following relation
\begin{align}
S_{\mr{HWZ}}(\mc{C}_{v}) = S_{\mr{Wall}}(\mc{C}_{v}) = S_{\mr{IW}}(\mc{C}_{v}) = S_{\mr{Wald}} \qquad \text{in a stationary regime}. \label{entropy relation for stationary spacetime}
\end{align}

So far, we have reviewed general properties of the HWZ entropy.
Now, let us focus on the case where the diffeomorphism invariant action is taken to be the Einstein--Hilbert action.
In this case, the HWZ entropy is given by
\begin{align}
    S_{\rm{HWZ}}(\mathcal{C}_v)=\left(1-v \partial_v \right)S_{\rm{Area}}(\mathcal{C}_v)+ \mathcal{O}(\epsilon^2) \,, 
\end{align}
and therefore, the Wall entropy on a background Killing horizon cross-section $\mc{C}_v$ reduces to
\begin{align}
    S_\mr{Wall} (\mc{C}_v) = S_\mr{Area}(\mc{C}_v) + \mc{O}(\ep^2)\,,
\end{align}
where $S_{\text{Area}}$ is the areal entropy\footnote{
Although we did not specify a choice of a background Killing horizon $\mathcal{H}^{+}$ here, if it is chosen as the event horizon of the full metric, the areal entropy simply represents the Bekenstein--Hawking entropy $S_{\text{Area}} = A_{\text{event}}/4 G \hbar$.
We note that, even in this case, no information about the surface being an event horizon is required to define the HWZ and Wall entropies. Therefore they are locally defined.}
defined as follows:
for a given codimension-2 surface $\mathcal{C}$ characterized by $(u,v) = (u(\varphi^{A}), v(\varphi^{A}))$ in the affinely parametrized Gaussian null coordinates, we define
\begin{align}
S_{\text{Area}}(\mathcal{C}) \coloneqq \frac{1}{4 G \hbar} A(\mathcal{C}) = \frac{1}{4 G \hbar} \int_{\mathcal{C}} \sqrt{\gamma_{{}_\mathcal{C}}} ~d \varphi^{1} \cdots d \varphi^{D-2}\,.
\label{def areal entropy}
\end{align}
Here, $\gamma_{{}_\mathcal{C}}$ is the determinant of the induced metric $\gamma_{{}_\mathcal{C}}{}_{ab}$ on $\mathcal{C}$.
Interestingly, it was shown that the HWZ entropy for the Einstein--Hilbert action relates with the areal entropy on the apparent horizon $\mc{T}_{v}$ on $\overline{\mc{N}}_{v}$,
\begin{align}
S_{\text{HWZ}}(\mathcal{C}_{v}) = S_{\text{Area}}(\mathcal{T}_{v}) + \mc{O}(\epsilon^2) = \frac{A_{\text{app}}}{4 G \hbar} + \mc{O}(\epsilon^2)\,. \label{SHWZ = Aapp in GR}
\end{align}
Although, this relation was derived for the Einstein--Hilbert action, it can be expressed as 
\begin{align}
    S_\mr{HWZ}(\mc{C}_v) = S_\mr{Wall}(\mc{T}_v) +\mc{O}(\ep^2)\,.
\end{align}
The key observation is that now the Wall entropy is defined for a general diffeomorphism invariant action. In the next section, we demonstrate that this relation indeed holds for any such action by identifying $\mc{T}_{v}$ as a generalized concept of the apparent horizon, which we call the \textit{entropic marginally outer trapped surface}.

\section{Entropy on the Entropic Marginally Outer Trapped Surface\label{sec EMOTS}}
In this section, we introduce the notion of an \textit{entropic expansion}, which is defined in association with a given entropy formula.
Based on this, we define an \textit{entropic marginally trapped surface} characterized by the vanishing of the entropic expansion.
We then show that the HWZ entropy can be interpreted as the Wall entropy evaluated on the entropic marginally outer trapped surface. These results generalize the relation~\eqref{SHWZ = Aapp in GR}, which is valid for the Einstein--Hilbert action, to a general diffeomorphism invariant gravitational action.

\subsection{Entropy for the Einstein--Hilbert action and the Expansion}
To proceed with the definition of the entropic expansion for a general diffeomorphism invariant gravitational action, let us see the relation between the expansion $\vartheta$ and the area of a codimension-2 surface.
As explained in the previous section, the Wall entropy associated with the Einstein--Hilbert action is the areal entropy given by Eq.~\eqref{def areal entropy}. Since $\sqrt{\gamma_{{}_\mathcal{C}}}/4G\hbar$ can be regarded as the entropy density associated with the areal entropy, let us express it as
\begin{align}
s^{\text{Area}}\coloneqq \frac{1}{4G\hbar} \sqrt{\gamma_{{}_\mathcal{C}}}\,. \label{sArea}
\end{align}

Let $n^{a}_{\pm}$ be the future-directed null normal vectors to $\mathcal{C}$, normalized such that $n^a_+ n_{-}{}_a=-\frac{1}{\Omega^2}$. Then, the expansion in the outward direction is defined by 
\begin{align}
    \vartheta_{+} \coloneqq \gamma_{{}_\mathcal{C}}^{a}{}_{b} \nabla_{a} n^{b}_{+}\,.
\end{align}
This expression is intrinsic to $\mathcal{C}$.
At this stage, $n_{\pm}^a$ are defined only on $\mathcal{C}$. Let us extend the definition of $n_{\pm}^a$ beyond the surface $\mathcal{C}$. We require $n_{+}^a$ to be tangent to affinely parametrized null geodesics, thus,
\begin{align}
    n_{+}^{b} \nabla_{b} n_{+}^{a} = 0\,,
\end{align}
and $n_{+}^a,\, n_{-}^a$ form a null tetrad, and the metric can be decomposed as
\begin{align}
g_{ab} = - \Omega^2 (n_{+}{}_{a} n_{-}{}_{b} + n_{+}{}_{b} n_{-}{}_{a}) + \gamma_{{}_\mathcal{C}}{}_{ab}\,.  \label{eq:null tetrad}
\end{align}
In particular, Eq.~\eqref{eq:null tetrad} indicates $n_{-}^{a} \nabla_{a} (n_{+}{}_{b} n_{+}^{b})=0$.
Using these extensions, the expansion can be expressed as
\begin{align}
    \vartheta_{+} &= \left( \delta^{a}{}_{b} + \Omega^2 n_{+}^{a} n_{-}{}_{b} + \Omega^2 n_{-}^{a} n_{+}{}_{b} \right)\nabla_{a} n_{+}^{b} \notag\\
 &= \nabla_{a} n^{a}_{+} + \Omega^2 n_{-}{}_{b} (n_{+}^{a} \nabla_{a} n_{+}^{b}) + \frac{1}{2} \Omega^2 n_{-}^{a} \nabla_{a} (n_{+}{}_{b} n_{+}^{b}) 
 \notag\\
 &= \nabla_{a} n_{+}^{a} \notag\\
 &= \frac{1}{\sqrt{\gamma_{{}_\mathcal{C}}}} \partial_{\mu} \left( \sqrt{\gamma_{{}_\mathcal{C}}} n^{\mu}_{+} \right) \notag\\
 &= n_{+}^{\mu} \partial_{\mu} \log \sqrt{\gamma_{{}_\mathcal{C}}} + \partial_{\mu} n^{\mu}_{+}\,.
 \end{align}
 Thus, using the notion $\sqrt{\gamma_{{}_\mathcal{C}}}/4G\hbar = s^{\text{Area}}$, we obtain the following expression:
 \begin{align}
     \vartheta_{+} = \frac{1}{s^{\text{Area}}} \partial_{\mu} \left( s^{\text{Area}} n_{+}^{\mu}\right) = n_{+}^{\mu} \partial_{\mu} \log s^{\text{Area}} + \partial_{\mu} n^{\mu}_{+}\,.
 \end{align}
 Motivated by this expression, we will define an \textit{entropic expansion} in the next subsection. 

\subsection{Entropic Expansions\label{subsub entropy}}
In this and the next subsections, we provide a discussion without referring to a specific definition of entropy, which is therefore applicable to any entropy formula. Here, we only assume that a concept of entropy, which satisfies the following requirements, is defined on any compact codimension-2 surface $\mathcal{C}$ which is parametrized by $\set{\varphi^{A}}$. 
We require that the entropy of a surface $\mathcal{C}$ is represented as the integral of an entropy $(\mathrm{D}-2)$-form $\bm{s}$, which can be written in terms of the entropy density $s$ as 
\begin{align}
 \bm{s} = s~d\varphi^1 \wedge \cdots \wedge d \varphi^{\mathrm{D}-2}\,.
\end{align}
We treat the entropy density $s$ as a scalar density\footnote{In many cases, for gravitational entropies such as the Bekenstein--Hawking entropy, the entropy density can be defined only after fixing a surface with respect to which it will be evaluated.
In our treatment, however, we first fix an expression of the entropy density $s$ and simply treat it as a scalar density field. Therefore, the entropy evaluated in our approach is meaningful only when it is computed on surfaces appropriately associated with the construction of the chosen form of $s$. We will revisit this point in Sec.~\ref{sec: application} with an application to the HWZ entropy.}
defined over the spacetime.
The entropy of a surface $\mathcal{C}$ can be expressed as 
\begin{align}
    S(\mathcal{C}) \coloneqq \int_{\mathcal{C}} \bm{s} = \int_{\mathcal{C}} s~d \varphi^{1} \cdots d \varphi^{D-2}\,.
\end{align}

We define an \textit{entropic expansion along $w^{a}$ direction} by
\begin{align}
    \Theta_{w} \coloneqq \frac{1}{s} \partial_{\mu}(w^{\mu} s ) = w^{\mu} \partial_{\mu}\log s + \partial_{\mu} w^{\mu} \,.
    \label{generalized expansion}
\end{align}
Note that this quantity can be expressed in terms of the Lie derivative, which is discussed in App.~\ref{app:LieD}. 

Let us summarize a few properties of $\Theta_{w}$:
\begin{itemize}
    \item \textbf{Linearity:} Entropic expansions are linear in $w^{a}$, thus, for two vectors $w_{1}^{a}$ and $w_{2}^{b}$, and for two constants $c_{1}$, $c_{2}$, an entropic expansion satisfies
\begin{align}
\Theta_{c_{1} w_{1} + c_{2} w_{2}} = c_{1} \Theta_{w_{1}} + c_{2} \Theta_{w_{2}}\,.
\end{align}
\item \textbf{Commutativity:}
For any vector fields $w_{1}^{a}$ and $w_{2}^{a}$,
\begin{align}
 \mathsterling_{w_{1}} \Theta_{w_{2}} - \mathsterling_{w_{2}} \Theta_{w_{1}} = \Theta_{[w_{1}, w_{2}]}\,.
\end{align}
In particular, when $[w_{1}, w_{2}]^{a} = 0$, we obtain
\begin{align}
 \mathsterling_{w_{1}} \Theta_{w_{2}} - \mathsterling_{w_{2}} \Theta_{w_{1}} = 0\,.
\end{align}
For example, for the coordinate basis $k^{a} = (\partial_{v})^{a}$ and $\ell^{a} = (\partial_{u})^{a}$, we obtain
\begin{align}
    \partial_{u} \Theta_{k} = \partial_{v} \Theta_{\ell}\,. \label{eq:commute kl}
\end{align}
\item \textbf{Covariance:}
One can show that the entropic expansion $\Theta_{w}$ is scalar under the coordinate transformations, notifying that $s$ is a scalar density. See App.~\ref{app:LieD}.
\item 
\textbf{Expressions in divergence free coordinates:}
If the components of the vector $w^{a}$ is divergence free with respect to the partial derivatives associated with the coordinates used there, $\partial_{\mu} w^{\mu} = 0$, the entropic expansion is evaluated as $\Theta_{w} = w^{\mu} \partial_{\mu} \log s$. In particular, the entropic expansion for the coordinate basis, $\ell^{a} = (\partial_{u})^{a}$ and $k^{a} = (\partial_{v})^{a}$ reduce to
\begin{align}
 \Theta_{\ell} = \partial_{u} \log s \,, \qquad   \Theta_{k} = \partial_{v} \log s\,.
\end{align}
As a result, the commutativity \eqref{eq:commute kl} follows directly from the commutativity of partial derivatives: $\partial_{u} \Theta_{k} = \partial_{u} \partial_{v} \log s = \partial_{v} \partial_{u} \log s = \partial_{v} \Theta_{l}$.
\end{itemize}

\subsection{Entropic Trapped Surfaces and Entropic Apparent Horizons\label{subsec: EMOTS}}
We define concepts of various trapped surfaces associated with the entropic expansion. Given a spacelike, compact, and boundaryless codimension-2 surface $\mathcal{C}$, one can introduce two classes of null vector fields orthogonal to $\mathcal{C}$, up to overall normalization. Let $n_{+}^{a}$ and $n_{-}^{a}$ be future-pointing, outward and inward null vector fields normal to $\mathcal{C}$, respectively.
As in the case of the Bekenstein--Hawking entropy, we extend the definition of $n_{+}^{a}$ from $\mathcal{C}$ to its neighborhood by requiring that
\begin{align}
n_{+}^{a}(\nabla_{a} n_{+}^{b}) n_{-}{}_{b} &= 0\,, \label{eq:n+ condition 1}\\
n_{-}^{a}(\nabla_{a} n_{+}^{b}) n_{+}{}_{b} &= 0\,, \label{eq:n+ condition 2}
\end{align}
which are automatically satisfied if $n_{+}^{a}$ is extended as an affinely parametrized null geodesic tangent vector along the $n_{+}^{a}$ direction and as a null vector field along the $n_{-}^{a}$ direction.

We define the outward entropic expansion of a surface $\mathcal{C}$ as
\begin{align}
    \Theta_{+} \coloneqq \Theta_{n_{+}} = \frac{1}{s} \partial_{\mu} \left(s n_{+}^{\mu}\right)\label{eq:def Theta+}
\end{align}
with $n_{+}$ extended beyond $\mathcal{C}$ as given in Eqs.~\eqref{eq:n+ condition 1} and \eqref{eq:n+ condition 2}.
We can express $\Theta_{+}$ in a manifestly intrinsic manner as follows:
First, let us introduce a scalar quantity $\tilde{s}$ by $\tilde{s} \coloneqq s/\sqrt{\gamma_{{}_\mathcal{C}}}$. Then, the outward entropic expansion can be expressed as 
\begin{align}
    \Theta_{+} &= \frac{1}{\tilde{s}}
    \frac{1}{\sqrt{\gamma_{{}_\mathcal{C}}}}\partial_{\mu} \left( \sqrt{\gamma_{{}_\mathcal{C}}} \tilde{s} n_{+}^{\mu}\right) \notag\\
    & = \frac{1}{\tilde{s}} \nabla_{\mu} \left( \tilde{s} n_{+}^{\mu} \right) \notag\\
    &= \nabla_{\mu} n_{+}^{\mu} + n_{+}^{\mu} \nabla_{\mu} \log \tilde{s}\,.
\end{align}
Since the first term is nothing but the usual expansion $\vartheta$ under the conditions~\eqref{eq:n+ condition 1} and \eqref{eq:n+ condition 2}, we obtain
\begin{align}
    \Theta_{+} &= \gamma_{{}_\mathcal{C}}^{\mu}{}_{\nu} \nabla_{\mu} n_{+}^{\nu} + n_{+}^{\mu} \nabla_{\mu} \log \tilde{s} \label{eq:Theta+ intrinsic way}\\
    &= \vartheta_{+} + n_{+}^{\mu} \nabla_{\mu} \log \tilde{s}\,. \label{eq:Theta+ theta+ with corrections}
\end{align}
The expression \eqref{eq:Theta+ intrinsic way} is manifestly independent of the extension of $n_{+}^{a}$ beyond $\mathcal{C}$. Furthermore, the expression \eqref{eq:Theta+ theta+ with corrections} clarifies the difference between the usual expansion and our entropic expansion. The inward entropic expansion $\Theta_{-}$ is defined in a similar way.

We define the concepts of trapped surfaces using the outward and inward entropic expansions on $\mathcal{C}$.
Specifically, we define the surface $\mathcal{C}$ to be:
\begin{itemize}
\item an \textit{Entropic Trapped Surface} if $\Theta_{+} \Theta_{-} > 0$,
\item an \textit{Entropic Marginally Outer Trapped Surface (E-MOTS)}, or equivalently an \textit{Entropic Apparent Horizon}, if $\Theta_{+} = 0$.
\end{itemize}
We further use the labels \textit{future}/\textit{past} to indicate that $\Theta_{-} < 0$ / $\Theta_{-} > 0$, respectively, on $\mathcal{C}$.
Note that by a future/past E-MOTS, we indicate $\Theta_{-} \neq 0$.

A marginally trapped surface is often referred to as an \textit{apparent horizon}, and accordingly, we refer to our E-MOTS as the \textit{Entropic Apparent Horizon}. However, throughout this paper, we primarily use the term E-MOTS\footnote{Note that, from the expression \eqref{eq:Theta+ theta+ with corrections}, whether the entropic apparent horizon lies outside or inside the apparent horizon can be determined by the sign of the second term of Eq.~\eqref{eq:Theta+ theta+ with corrections}, which depends on the action.}.

Various notions of trapped surfaces in general relativity can be extended to their entropic counterparts. For example, we can define a \textit{null strictly stable E-MOTS} as a surface satisfying $\Theta_{+}=0$ and $\mathsterling_{n_{-}}\Theta_{n_{+}} \neq 0$.
We will see later that future/past E-MOTSs in a stationary spacetime coincide with a null strictly stable E-MOTS.
We also use the term \textit{entropic (marginally outer) trapped tube} to denote a codimension-1 hypersurface foliated by entropic (marginally outer) trapped surfaces.  

\subsection{Entropy on Entropic Marginally Outer Trapped Surfaces}
In the following, we once again consider first-order perturbations around a stationary black hole as in Sec.~\ref{sec: review on HWZ}.  

We consider an entropy density depending on the perturbation parameter $\epsilon$ in general, and express it as 
\begin{align}
s(\epsilon; u,v, \varphi^{A}) = \bg{s}(u,v,\varphi^{A}) + \epsilon \delta s(u,v,\varphi^{A}) + \mathcal{O}(\epsilon^2)\,.
\end{align}
Accordingly, the entropy evaluated on a codimension-2 surface $\mathcal{C}$ depends on $\epsilon$, and is given by
\begin{align}
S(\epsilon; \mathcal{C}) = \int_{\mathcal{C}} s(\epsilon; u(\varphi^{A}), v(\varphi^{A}), \varphi^{A}) d^{\mathrm{D}-2} \varphi\,.
\end{align}
Note that in Sec.~\ref{sec: review on HWZ}, the dependence on $\epsilon$ was omitted for notational simplicity. According to the notation adopted in this section, where the $\epsilon$-dependence is made explicit, a quantity written as $S_{\text{HWZ}}(\mc{C}_{v})$ in Sec.~\ref{sec: review on HWZ} should now be written as $S_{\text{HWZ}}(\epsilon;\mc{C}_{v})$.

The entropic expansion for the full metric $g_{ab}(\epsilon)$ is similarly defined by
\begin{align}
\Theta_{w}(\epsilon; u,v,\varphi^{A}) = w^{\mu} \partial_{\mu} \log s(\epsilon;u,v,\varphi^{A}) + \partial_{\mu} w^{\mu}\,.
\end{align}
The perturbations of the entropic expansion can be expanded as
\begin{align}
\Theta_{w}(\epsilon; u,v,\varphi^{A}) = \,\bg{\Theta}_{w}(u,v,\varphi^{A}) + \epsilon \delta \Theta_{w}(u,v,\varphi^{A}) + \mathcal{O}(\epsilon^2)\,,
\end{align}
where
\begin{align}
 \bg{\Theta}_{w} &= w^{\mu} \partial_{\mu} \log \bg{s}\,, \\
 \delta \Theta_{w} &= w^{\mu} \partial_{\mu} \left( \frac{\delta s}{\bg{s}} \right)\,.
\end{align}

Our main statement in this paper is the following:
\\
~ \\
\begin{samepage}
\begin{itshape}Let $\mathcal{H}^{+}$ be a background future or past entropic marginally outer trapped tube, foliated by background entropic marginally trapped surfaces $\mc{C}_{v}$, with respect to the background metric $\stackrel{\circ}{g}_{ab}$, with the outward and inward null normal vector $k^{a}$ and $\ell^{a}$, respectively.
Then, for a given fixed value of $v$, there exists an entropic marginally outer trapped surface $\mathcal{T}_{v}$ on $\overline{\mathcal{N}}_{v}$ for the perturbed metric at first-order. Furthermore, the entropy evaluated on $\mathcal{T}_{v}$ satisfies
\begin{align}
S(\epsilon; \mathcal{T}_{v}) = (1 - v \partial_{v}) S(\epsilon;\mathcal{C}_{v}) + \mathcal{O}(\epsilon^2 )\,. \label{ST = 1 -v del v SCv}
\end{align}
\end{itshape}
\end{samepage}
\\
~

Let us proceed to prove this statement.
Since $\mathcal{C}_{v}$ is a future or past entropic marginally outer trapped surface for the background, we obtain
\begin{align}
\stackrel{\circ}{\Theta}_{+} &=\,
\stackrel{\circ}{\Theta}_{k}\!(0, v, \varphi^{A}) = \partial_{v} \log \stackrel{\circ}{s} (0, v, \varphi^{A}) = 0\,, \label{eq:Theta+ bg}\\
\stackrel{\circ}{\Theta}_{-}
&=\,
\stackrel{\circ}{\Theta}_{\ell}(0, v, \varphi^{A}) =\, \partial_{u} \log \stackrel{\circ}{s}\!(0, v, \varphi^{A}) \neq 0\,.
\label{eq:Theta- bg}
\end{align}
Recall that $\mathcal{C}_{v}$ corresponds to the surface $u = 0$ and $v = \text{const.}$ in the affinely parametrized Gaussian null coordinates.
Eq.~\eqref{eq:Theta+ bg} implies that
\begin{align}
\mathsterling_{\xi} \!\bg{s}~|_{\mathcal{H}^+} = \kappa v \partial_{v}\! \bg{s} (0,v,\varphi^{A}) = 0\,.
\end{align}
Hence, the boost weight argument ensures that $\stackrel{\circ}{s}$ can be expressed as 
\begin{align}
\stackrel{\circ}{s} = \bg{s}_{0}(\varphi^{A}) + \bg{s}_{1}(\varphi^{A}) u v + \mathcal{O}((uv)^2)\,,  \label{sbg = s0 + s1 uv}
\end{align}
which results in
\begin{align}
\stackrel{\circ}{\Theta}_{\ell}\!(0,v,\varphi^{A}) = C(\varphi^{A}) v\,,
\end{align}
where $C(\varphi^{A}) = \bg{s}_{1}(\varphi^{A})/\bg{s}_{0}(\varphi^{A})$.
Then, from the commutativity of the entropic expansion with respect to the coordinate basis (Eq.~\eqref{eq:commute kl}), we find
\begin{align}
 \partial_{u}\! \stackrel{\circ}{\Theta}_{k} {} = \partial_{v}\! \stackrel{\circ}{\Theta}_{\ell} \!{} \stackrel{\mathcal{H}^{+}}{=} \partial_{v} \left( C(\varphi^{A}) v \right) = C(\varphi^{A})\,.
\label{S(T) by Thetal}
\end{align}
Thus, we obtain
\begin{align}
\stackrel{\circ}{\Theta}_{\ell}\!(0,v, \varphi^{A}) = \partial_{u} \!\stackrel{\circ}{\Theta}_{k} \!(0,v,\varphi^{A}) v \,. \label{eq:Thetal bg = pdu Thetak bg v}
\end{align}
We will use this expression below when evaluating the entropy formula\footnote{
Through Eq.~\eqref{eq:Thetal bg = pdu Thetak bg v}, the condition $\bg{\Theta}_{-} \neq 0$ can be expressed as $\mathsterling_{\bm{n}_{-}} \!\!\bg{\Theta}_{+} = \partial_{u} \!\bg{\Theta}_{k} \neq 0$ as well. Thus, our requirement for $\mathcal{C}_{v}$ to be a future/past E-MOTS can be equivalent to require that it is a null stable E-MOTS. In fact, in the example of the Einstein--Hilbert action studied by Hollands--Wald--Zhang~\cite{Hollands:2024vbe}, $\mc{C}_{v}$ is assumed to be a null stable MOTS.
}.

Let us consider a spacelike, compact, and boundaryless codimension-2 surface $\mathcal{T}_{v_{0}}(\epsilon)$ on $\overline{\mathcal{N}}_{v_{0}}$, which can be specified by the coordinates $u = u_{\mathcal{T}_{v_{0}}}(\epsilon; \varphi^{A})$, $v = v_{0}$.
Assuming that $\mathcal{T}_{v_{0}}(0) = \mathcal{C}_{v_{0}}$, 
we can expand $u_{\mathcal{T}_{v_{0}}}(\epsilon; \varphi^A)$ as $u_{\mathcal{T}_{v_{0}}}(\epsilon; \varphi^{A}) = \delta u_{\mathcal{T}_{v_{0}}} (\varphi^{A}) \epsilon + \mathcal{O}(\epsilon^2)$.
Let $n_{+}^{a}(\epsilon)$ be an outward null normal vector field of $\mathcal{T}_{v_{0}}(\epsilon)$, which satisfies $n_{+}^{\mu}(0; \varphi^{A}) = k^{\mu}$ in the affinely parametrized Gaussian null coordinates. 
We then expand $n_{+}^{a}(\epsilon)$ in $\epsilon$ as 
\begin{align}
n_{+}^{\mu}(\epsilon;\varphi^{A}) = k^{\mu} + \epsilon \delta n_{+}^{\mu}(\varphi^{A}) + \mathcal{O}(\epsilon^2)\,,
\end{align}
in the affinely parametrized Gaussian null coordinates. Note that $k^{\mu} = \delta^{\mu}_{v}$ is a constant.
We also introduce the inward null normal of $\mathcal{T}_{v_{0}}(\epsilon)$ as
\begin{align}
    n_{-}^{\mu} = \ell^{\mu}\,,
\end{align}
which is taken to be independent of $\epsilon$.

Here, we summarize the constraints imposed on $\delta n_{+}^{a}$:
\begin{itemize}
    \item \textbf{Nullness of $n_{+}^{a}$:} Since the norm of $n_{+}^a$ can be evaluated as
\begin{align}
    g_{\mu\nu} n_{+}^{\mu} n_{+}^{\nu}|_{\mathcal{T}} &= 2 \epsilon \bg{g}_{\mu\nu}(0,v_{0},\varphi^{A}) k^{\mu} \delta n^{\nu}_{+}(\varphi^{A}) + \mathcal{O}(\epsilon^2) \notag\\
    &=- 2 \epsilon \delta n_{+}^{u}(\varphi^{A}) + \mathcal{O}(\epsilon^2)\,.
\end{align}
The requirement of the nullness of $n_{+}^a$ can be expressed as 
\begin{align}
    \delta n_{+}^{u}(\varphi^{A}) = 0\,. \label{eq:delta nu = 0}
\end{align}
    \item \textbf{Ambiguity of the overall factor:}
    Since the normalization of $n_{+}^{a}$ is unfixed, we can replace $n_{+}^{a} \to F(\epsilon;\varphi^{A}) n_{+}^{a}$ with a positive function $F$ satisfying $\bg{F} = 1$, instead of $n_{+}^{a}$. In other words, $\delta n_{+}^{a}$ defined here has an ambiguity
\begin{align}
\delta n_{+}^{a} \rightarrow  \delta n_{+}^{a} + \delta F(\varphi^{A}) k^{a}\,.
\end{align}
In index notation, it means $\delta n_{+}^{v} \rightarrow \delta n_{+}^{v} + \delta F(\varphi^{A})$.
\end{itemize}

So far, $n_{+}^{a}(\epsilon)$ has only been defined on $\mathcal{T}_{v_{0}}(\epsilon)$. In order to use a non-intrinsic definition of $\Theta_{+}$ given in Eq.~\eqref{eq:def Theta+}, we need to extend $n_{+}^{a}$ to a vector field defined in a neighborhood of $\mathcal{T}_{v_{0}}$, i.e., promote it as $n_{+}^{\mu}(\epsilon;\varphi^{A}) \to n_{+}^{\mu}(\epsilon;u,v,\varphi^{A})$. This extension must satisfy the following three conditions:
\begin{enumerate}
    \item \textbf{$\bm{n_{+}^{\mu}(\epsilon;u_{\mathcal{T}_{v_{0}}},v_{0},\varphi^{A}) = n_{+}^{\mu}(\epsilon,\varphi^{A})}$:} Expanding in $\epsilon$, this condition can be simply expressed as 
    \begin{align}
        \delta n_{+}^{\mu}(0, v_{0}, \varphi^{A}) = \delta n_{+}^{\mu}(\varphi^{A})\,.
    \end{align}
    \item \textbf{Eq.~\eqref{eq:n+ condition 1}, $\bm{\left.n_{-}^{a}(\nabla_{a} n_{+}^{b}) n_{+}{}_{b}\right|_{\mathcal{T}_{v_{0}}} = 0}$:} 
    The left-hand side of this condition can be evaluated as
    \begin{align}
    &n_{-}^{a}(\nabla_{a} n_{+}^{b}) n_{+}{}_{b}|_{\mathcal{T}_{v_{0}}} \notag\\
     & = - \epsilon \left(
        \partial_{u}\delta n_{+}^{u} (0,v_{0}, \varphi^{A})
        + \bg{\alpha}(0,v_{0},\varphi^{A}) \delta u_{\mathcal{T}_{v_{0}}}(\varphi^{A}) + \bg{\beta}_{A}(0,v_{0},\varphi^{A}) \delta n^{A}(\varphi^{A})
        \right) + \mathcal{O}(\epsilon^2)\,.
    \end{align}
    Thus, we obtain
    \begin{align}
    \partial_{u}\delta n_{+}^{u} (0,v_{0}, \varphi^{A})
        = - \bg{\alpha}(0,v_{0},\varphi^{A}) \delta u_{\mathcal{T}_{v_{0}}}(\varphi^{A}) - \bg{\beta}_{A}(0,v_{0},\varphi^{A}) \delta n^{A}(\varphi^{A})\,. \label{eq:pdu deltanu}
    \end{align}
    \item \textbf{Eq.~\eqref{eq:n+ condition 2}, $\bm{\left.n_{+}^{a}(\nabla_{a} n_{+}^{b}) n_{-}{}_{b}\right|_{\mathcal{T}_{v_{0}}}=0}$:} In a similar manner, we obtain
    \begin{align}
     \partial_{v}\delta n_{+}^{v} (0,v_{0}, \varphi^{A})
        = ~ \bg{\alpha}(0,v_{0},\varphi^{A}) \delta u_{\mathcal{T}_{v_{0}}}(\varphi^{A}) + \bg{\beta}_{A}(0,v_{0},\varphi^{A}) \delta n^{A}(\varphi^{A})\,. \label{eq:pdv deltanv}
    \end{align}
\end{enumerate}
Since the right-hand sides of Eqs.~\eqref{eq:pdu deltanu} and $\eqref{eq:pdv deltanv}$ are opposite in sign, we find
\begin{align}
  \partial_{u}\delta n_{+}^{u} (0,v_{0}, \varphi^{A}) +  \partial_{v}\delta n_{+}^{v} (0,v_{0}, \varphi^{A}) = 0\,. \label{eq:div deltan uv = 0}
\end{align}

Now, we are ready to prove our statement.
In the following, we omit the arguments $v_{0}$ and $\varphi^{A}$ if they are clear from the context.
The first step is to expand the outward entropic expansion on $\mathcal{T}_{v_{0}}$ in $\epsilon$,
\begin{align}
\Theta_{+}(\epsilon) &= 
\Theta_{n_{+}}(\epsilon; u_{\mathcal{T}_{v_{0}}}) \notag\\
&= \Theta_{k}(\epsilon; u_{\mathcal{T}_{v_{0}}}) + \epsilon \Theta_{\delta n_{+}}(\epsilon; u_{\mathcal{T}_{v_{0}}}) + \mathcal{O}(\epsilon^2) \notag\\ 
&= 
\bg{\Theta}_{k}(0) + \epsilon \delta \Theta_{k}(0) + \epsilon \partial_{u} \bg{\Theta}_{k}(0) \delta u_{\mathcal{T}_{v_{0}}} + \epsilon \bg{\Theta}_{\delta n_{+}}(0) + \mathcal{O}(\epsilon^2) \notag\\
& =
\epsilon \left( \delta \Theta_{k}(0) + \frac{1}{v_{0}} \bg{\Theta}_{\ell}(0) \delta u_{\mathcal{T}_{v_{0}}} + \bg{\Theta}_{\delta n_{+}}(0) \right) + \mathcal{O}(\epsilon^2)\,.
\end{align}
Here, we use Eqs.~\eqref{eq:Theta+ bg} and \eqref{eq:Thetal bg = pdu Thetak bg v} in the last equality. Since Eq.~\eqref{eq:Theta- bg} ensures that the coefficient of $\delta u_{\mathcal{T}}$ does not vanish, we can choose the function $\delta u_{\mathcal{T}_{v_{0}}}$ as 
\begin{align}
\delta u_{\mathcal{T}_{v_{0}}} = - v_{0} \frac{\delta \Theta_{k}(0) + \bg{\Theta}_{\delta n_{+}}(0)}{\bg{\Theta}_{\ell}(0)}\,, \label{eq:deltau EMOTS}
\end{align}
and with this choice, the outward entropic expansion on $\mathcal{T}_{v_{0}}$ vanishes up to order $\epsilon^2$:
\begin{align}
\Theta_{+}(\epsilon) = \mathcal{O}(\epsilon^2)\,.
\end{align}
Thus, with the choice in Eq.~\eqref{eq:deltau EMOTS}, the surface $\mathcal{T}_{v_{0}}(\epsilon)$ is an entropic marginally outer trapped surface for the perturbed metric $g_{ab}(\epsilon)$, which provides the proof of our first statement.

Next, let us evaluate the entropy on the entropic marginally outer trapped surface $\mathcal{T}_{v_{0}}(\epsilon)$ constructed above.
The entropy on $\mathcal{T}_{v_{0}}(\epsilon)$ can be evaluated as
\begin{align}
     S(\epsilon; \mathcal{T}_{v_{0}}(\epsilon)) 
     & = \int s(\epsilon; u_{\mathcal{T}}) d^{\mathrm{D}-2} \varphi \notag\\
    &= \int \left[ s(\epsilon;0) + \epsilon \partial_{u}\! \stackrel{\circ}{s}\!(0) \delta u_{\mathcal{T}_{v_{0}}}  + \mathcal{O}(\epsilon^2) \right]d^{\mathrm{D}-2} \varphi \notag\\
    &=S(\epsilon; \mathcal{T}_{v_{0}}(0)) + \epsilon \int \left[\stackrel{\circ}{s}\!(0) \stackrel{\circ}{\Theta}_\ell\!(0) \delta u_{\mathcal{T}_{v_{0}}} \right] d^{\mathrm{D}-2} \varphi +\mathcal{O}(\epsilon^2) \,.
\end{align}
Using $\mathcal{T}_{v_{0}}(0)=\mathcal{C}_{v_{0}}$ and substituting the expression for $\delta u_{\mathcal{T}}$ given in Eq.~\eqref{eq:deltau EMOTS}, we obtain
\begin{align}
S(\epsilon; \mathcal{T}_{v_{0}}(\epsilon))
= S(\epsilon; \mathcal{C}_{v_{0}}) - v_{0}
\int \left[
 \epsilon \stackrel{\circ}{s}\!(0) \delta \Theta_{k}(0) + \epsilon \stackrel{\circ}{s}\!(0) \bg{\Theta}_{\delta n_{+}}(0)
\right] d^{\mathrm{D}-2} \varphi + \mathcal{O}(\epsilon^2)\,. \label{S(T)=S-vint}
\end{align} 

Let us evaluate the first term of the integral in Eq.~\eqref{S(T)=S-vint}. Since $\partial_{v} \!\stackrel{\circ}{s}\!\!(0) = 0$, 
we obtain $\partial_{v} s(\epsilon; 0) = \mathcal{O}(\epsilon)$. Therefore,
\begin{align}
 \Theta_{k}(\epsilon;0) = \frac{\partial_{v} s(\epsilon;0)}{\stackrel{\circ}{s}\!(0)} + \mathcal{O}(\epsilon^2)\,,
\end{align}
and thus,
\begin{align}
 \epsilon \delta \Theta_{k}(0) = \frac{\partial_{v} s (\epsilon;0)}{\stackrel{\circ}{s}\!(0)} + \mathcal{O}(\epsilon^2)\,.
\end{align}
Hence,
\begin{align}
\int \bg{s}(0) \epsilon \delta \Theta_{k}(0) d^{\mathrm{D}-2}\varphi = \partial_{v} \int s(\epsilon;0) d^{\mathrm{D}-2} \varphi + \mathcal{O}(\epsilon^2) = \partial_{v} S(\epsilon; \mathcal{C}_{v}) + \mathcal{O}(\epsilon^2)\,.
\end{align}

The second term of the integral in Eq.~\eqref{S(T)=S-vint} can be evaluated as follows.
By the definition of the entropic expansion, we find
\begin{align}
& \int \bg{s}(0) \epsilon \bg{\Theta}_{\delta n_{+}}(0) d^{\mathrm{D}-2}\varphi
= \epsilon \int \partial_{\mu}(\delta n_{+}^{\mu} \bg{s} (0)) d^{\mathrm{D-2}} \varphi \notag\\
&= \epsilon \int \left[ 
\left( \partial_{u} \delta n_{+}^{u}(0) + \partial_{v} \delta n_{+}^{v}(0) \right) \bg{s}(0) + \delta n_{+}^{u} \partial_{u} \bg{s}(0) + \delta n_{+}^{v} \partial_{v} \bg{s}(0)
+\partial_{A}(\delta n_{+}^{A} \bg{s} (0)) 
\right]d^{\mathrm{D-2}} \varphi\,.
\end{align}
Furthermore, $\partial_{u}\delta n_{+}^{u}(0) + \partial_{v} \delta n_{+}^{v}(0)$, $\delta n_{+}^{u}(0)$, and $\partial_{v} s(0)$ vanish from Eqs.~\eqref{eq:div deltan uv = 0}, \eqref{eq:delta nu = 0}, and \eqref{eq:Theta+ bg}, respectively. Therefore, the integrand reduces to a total derivative, and it must vanish because the surface $\mathcal{T}_{v_{0}}$ is assumed to be boundaryless, so we obtain
\begin{align}
\int \bg{s}(0) \epsilon \bg{\Theta}_{\delta n_{+}}(0) d^{\mathrm{D}-2}\varphi
= \epsilon \int \partial_{A}(\delta n_{+}^{A} \bg{s} (0)) 
d^{\mathrm{D-2}} \varphi = 0\,.
\end{align}

Summarizing the above calculations, we finally obtain the following result:
\begin{align}
S(\epsilon; \mathcal{T}_{v}(\epsilon))
&=  (1 - v \partial_{v}) S(\epsilon; \mathcal{C}_{v}) + \mathcal{O}(\epsilon^2)\,, 
\end{align}
where we have replaced $v_{0} \to v$.

Finally, let us comment on the dependence on the definition of the entropy density $s$ beyond the background Killing horizon $\mathcal{H}^{+}$. While many entropy formulae on $\mathcal{H}^{+}$ are well-established, the extension of such definitions away from $\mathcal{H}^{+}$ is still under active discussion. 
Our result can be applied to any given extension. However, the position of the E-MOTS depends on the choice of extension.
In our first-order analysis, all the extension dependence is encoded in the function $\bg{s}_{1}(\varphi^{A})$ appearing in the expression of the background entropy density \eqref{sbg = s0 + s1 uv}: The numerator of Eq.~\eqref{eq:deltau EMOTS} is determined by the quantities on $\mathcal{H}^{+}$, while the denominator is not and reduces to $\bg{s}_{1}$ as $\bg{\Theta}_{\ell}(0)/v_{0} = \bg{s}_{1}(\varphi^{A})$.

In the above analysis, we have considered a one-parameter family of metrics $g_{ab}(\epsilon)$ with $\epsilon \in [0,1]$. The result for the physical metric $g_{ab} = g_{ab}(1)$ is obtained simply by setting $\epsilon = 1$: 
\begin{align}
S(\mathcal{T}_{v})
&=  (1 - v \partial_{v}) S(\mathcal{C}_{v}) + \mathcal{O}(\text{second-order perturbations})\, .
\end{align}
Here, we use the notion $S(1; \mathcal{T}_{v}(1)) = S(\mathcal{T}_{v})$, and similarly for other quantities. It should be noted that the parameter $\epsilon$ serves to count the order of perturbations, and the terms written as $\mathcal{O}(\epsilon^2)$ include expressions such as $(\epsilon \delta g)^2$ and $\epsilon^2 \delta^2 g$, which are of the second-order in the perturbative expansion. The validity of the perturbative expansion is therefore ensured by the smallness of $\delta g$ and $\delta^2 g$ themselves.

\subsection{An Application to the Dynamical Entropy\label{sec: application}}
As we have seen in Sec.~\ref{subsub: Wall}, the HWZ entropy on $\mathcal{C}_{v}$ can be expressed by the Wall entropy by Eq.~\eqref{relation between HWZ and Wall}. By setting $\epsilon = 1$, we obtain 
\begin{align}
    S_\mr{HWZ}(\mc{C}_v) = (1-v\del_v )S_\mr{Wall}(\mc{C}_v) + \mathcal{O}(\text{second-order perturbations}) \,.
    \label{relation between HWZ and Wall again}
\end{align} 
The Wall entropy, as well as the HWZ entropy, is defined on the background Killing horizon $\mc{H}^{+}$ associated with the background metric $\bg{g}_{ab}$ so far. To apply the results in the previous section, we need an extension of the definition of $S_{\text{Wall}}$ from $\mc{H}^{+}$ to the neighborhood of it, and a $(\mr{D}-2)$-form field $\bm{s}^{\text{Wall}}$ compatible to the definition of $S_{\text{Wall}}$.
In the following, we provide an algorithm to construct the entropy $(\mr{D}-2)$-form $\bm{s}^{\text{Wall}}$ as a field in the spacetime.

First, let us consider a codimension-2 surface $\hat{\mc{C}}$ on $\overline{\mathcal{N}}_{v_{0}}$, characterized by $(u,v) = (u_{\hat{\mc{C}}}(\phiv),v_{0})$, with a perturbatively small $u_{\hat{\mc{C}}}(\phiv)$.
Then, we define the null hypersurface $\hat{\mc{N}}$ generated by the outward null geodesics orthogonal to $\hat{\mc{C}}$, with respect to the physical metric $g_{ab}$\footnote{
Generally, the null geodesics failed to form a null hypersurface when they intersect each others. Actually, what we need here is to define a null hypersurface only in a neighborhood of $\hat{\mathcal{C}}.$ }.
Since the null hypersurface $\hat{\mc{N}}$ differs from the background Killing horizon $\mc{H}^{+}$ by a perturbative amount, it can also be regarded as a background Killing horizon, as noted in Sec.~\ref{subsub: geometrical setup}.
Let $\bg{\hat{g}}_{ab}$ be the background stationary metric with the Killing horizon that can be identified with $\hat{\mc{N}}$. To distinguish the choice of background Killing horizons, we now denote the one associated with the background metric $\bg{g}_{ab}$, previously denoted as $\mathcal{H}^{+}$, by $\mathcal{H}^{+}\bigl(\bg{g}\bigr)$, and the one associated with $\bg{\hat{g}}$ by $\mathcal{H}^{+}\bigl( \bg{\hat{g}} \bigr)$. 
Since the construction of the Wall entropy, as well as the HWZ entropy, is applicable to any background Killing horizon, we can define the Wall entropy on $\hat{\mc{N}} \equiv \mathcal{H}^{+}\bigl( \bg{\hat{g}} \bigr)$ in the same way as on $\mathcal{H}^{+}\bigl( \bg{g} \bigr)$. Thus, by introducing the affinely parametrized Gaussian null coordinates $(\hat{u}, \hat{v}, \varphi^{A})$ associated with $\hat{\mc{N}} = \mathcal{H}^{+}\bigl( \bg{\hat{g}} \bigr)$ and a one-parameter family of metrics $\hat{g}_{ab}(\hat{\epsilon})$, 
the null hypersurface $\mathcal{H}^{+}\bigl( \bg{\hat{g}} \bigr)$ can be foliated by the $\hat{v}$ constant sections $\hat{\mathcal{C}}_{\hat{v}}$ and we can define the Wall entropy form on $\hat{\mc{C}}_{\hat{v}}$, denoted as $\bm{s}^{\text{Wall}}_{\hat{\mc{C}}_{\hat{v}}}$, as
\begin{align}
    \bm{s}^{\text{Wall}}_{\hat{\mc{C}}_{\hat{v}}} &\coloneqq 
    - \frac{8 \pi}{\hbar}
    \hat{\bm{\epv}}^{\hat{\mc{C}}_{\hat{v}}} {E}_{\hat{R}}^{\hat{u}\hat{v}\hat{u}\hat{v}} + \frac{4\pi}{\kappa \hbar} P_{\hat{\xi}}(\hat{g},\mathsterling_{\hat{\xi}} \hat{g}) \hat{\bm{\epv}}^{\hat{\mc{C}}_{\hat{v}}} \notag\\
&\qquad
    - \hat{v} \frac{2 \pi}{\hbar}
    \hat{\bm{\epv}}^{\hat{\mc{C}}_{\hat{v}}} \left(- 2 {E}_{\hat{R}}^{\hat{u}\hat{v}\hat{u}A}\hat{\beta}_A + 2 {W}^{\hat{u}\hat{v}\hat{u}}
    + 4{E}_{\hat{R}}^{\hat{u}AB\hat{v}} \hat{K}_{AB}
    + \displaystyle\sum_{w\ge 1} {T}_{(w-1)}(\hat{\nabla}\cdots\hat{\nabla} \hat{R})_{(-w)} \right) \notag\\
   & \qquad + \mathcal{O}(\hat{\epsilon}^2)
    \,, \label{swallhat}
\end{align}
where any hatted quantity is understood to be defined with respect to the one-parameter family of metrics $\hat{g}_{ab}(\hat{\epsilon})$ and the affinely parametrized Gaussian null coordinates with $\hat{\mc{N}}$.
Translation to the original perturbative expansion by $\epsilon$ can be done by identifying $\hat{\epsilon} \equiv \epsilon$ and expressing the background metric in the hatted gauge $\bg{\hat{g}}_{ab}$ by the original gauge, $\bg{\hat{g}}_{ab} = \bg{g}_{ab} + \epsilon \mathsterling_{\eta} \bg{g}_{ab} + {\mathcal{O}}(\epsilon^2)$, where $\eta^a$ is the generator of the diffeomorphism which connects two background metrics. 
Then, the extended Wall entropy on $\hat{\mathcal{C}}_{\hat{v}}$ is given by
\begin{align}
S_{\mr{Wall}}(\hat{\mc{C}}_{\hat{v}}) = \int_{\hat{\mathcal{C}}_{\hat{v}}} \bm{s}^{\mr{Wall}}_{\hat{\mathcal{C}}_{\hat{v}}}\,.
\end{align}

Now, for a given $\hat{\mc{C}}$, the Wall entropy form is defined as a field on $\hat{\mc{N}}$. We can then calculate the $\hat{v}$-derivative of the entropy density and the outward entropic expansion with respect to $\hat{\mathcal{C}}$. Therefore, we can look for E-MOTS on $\overline{\mathcal{N}}_{v_{0}}$ for the physical metric.

Assuming that an E-MOTS $\mc{T}_{v_{0}}$ for a fixed $\epsilon$ can be found by this method, we can then construct a foliation of a portion of $\overline{\mathcal{N}}_{v_{0}}$ that includes $\mc{T}_{v_{0}}$ and $\mathcal{C}_{v_{0}}$ as leaves, for example, we may consider a one-parameter family of surfaces $\hat{\mc{C}}^{\lambda}$ defined by $(u, v) = (\lambda u_{\mathcal{T}}(\varphi^{A}), v_{0})$ with $0 \leq \lambda \leq 1$. 
Now, repeating the above construction of the Wall entropy form starting from each $\hat{\mc{C}}^{\lambda}$, we can define the extension of a entropy density form $\bm{s}^{\mr{Wall}}$ over $\overline{\mathcal{N}}_{v_{0}}$ by
\begin{align}
\bm{s}^{\mr{Wall}} \stackrel{~\hat{\mathcal{C}}^{\lambda}}{\coloneqq} \bm{s}^{\text{Wall}}_{\hat{\mathcal{C}}^{\lambda}}\,.
\end{align}
In general, the Wall entropy on an arbitrary hypersurface $\hat{\mc{C}}$ does not coincide with the surface integral of our $s^{\mr{Wall}}$ unless the surface $\hat{\mc{C}}$ is a leaf of the foliation $\hat{\mathcal{C}}^{\lambda}$,
\begin{align}
S_{\text{Wall}}(\hat{\mc{C}}) \neq \int_{\hat{\mc{C}}} \bm{s}^{\mr{Wall}} \quad \text{when~}\ \hat{\mc{C}} \notin \{\hat{\mc{C}}^{\lambda}\}_{\lambda \in [0,1]} \,.
 \end{align}
However, since $\mathcal{C}_{v_{0}} = \hat{\mathcal{C}}^{\lambda = 0}$ and the E-MOTS $\mathcal{T}_{v_{0}} = \hat{\mathcal{C}}^{\lambda = 1}$ are leaves of the foliation by our construction, we obtain
\begin{align}
S_{\text{Wall}}(\mc{C}_{v_{0}}) &= \int_{\mc{C}_{v_{0}}} \bm{s}^{\text{Wall}}\,,  \\
S_{\text{Wall}}(\mc{T}_{v_{0}}) &= \int_{\mc{T}_{v_{0}}} \bm{s}^{\text{Wall}}\,.
\end{align}

The Wall entropy $(\mr{D} - 2)$-form on $\mathcal{H}^{+}\bigl(\bg{g}\bigr)$ is covariantly constructed from the metric, the curvature tensors, and their covariant derivatives.
Since the Killing vector $\xi$ for the background metric $\bg{g}_{ab}$ reduces to $\kappa v \partial_{v}$ on $\mathcal{H}^{+}\bigl(\bg{g}\bigr)$, we can confirm that
\begin{align}
\partial_{v}\! \bg{s}^{\,\text{Wall}} \stackrel{\mathcal{H}^{+}\bigl(\bg{g}~\bigr)}{=} 0\,,
\end{align}
and hence, the background Killing horizon $\mathcal{H}^{+}\bigl(\bg{g}\bigr)$ is also an E-MOTT for $\bg{g}_{ab}$.
Hence, provided that $\partial_{u}\bm{s}^{\text{Wall}}|_{\mathcal{H}^{+}\bigl(\bg{g}~\bigr)} \neq 0$, the extended Wall entropy satisfies our assumptions in the previous subsection. Thus, in this situation, we can apply the results in the previous section to the extended Wall entropy and obtain
\begin{align}
    S_{\mr{Wall}} (\mc{T}_{v}) = (1 - v \partial_{v}) S_{\mr{Wall}}(\mc{C}_{v}) + \mathcal{O}(\text{second-order perturbations})\,.
\end{align}
Then, combining Eq.~\eqref{relation between HWZ and Wall again}, we obtain
\begin{align}
    S_\mr{HWZ} (\mc{C}_{v}) = S_\mr{Wall}(\mc{T}_{v}) + \mc{O}(\text{second-order perturbations})\,.
    \label{main equation}
\end{align}
This is our main result.

\section{Summary and Discussion}
In this paper, we investigated the properties of the dynamical black hole entropy proposed by Hollands, Wald, and Zhang \cite{Hollands:2024vbe} for arbitrary diffeomorphism invariant gravitational action.
After reviewing the basic tools, such as the affinely parametrized Gaussian null coordinates and boost weight argument, as well as the basic definition of the dynamical black hole entropy in Sec.~\ref{sec: setup} and Sec.~\ref{sec: review on HWZ}, we proposed the definitions of the entropic expansion and the entropic marginally outer trapped surface (E-MOTS), given by Eqs.~\eqref{generalized expansion} and in Sec.~\ref{subsec: EMOTS}, respectively, associated with any given entropy density $s$.
When the entropy density $s$ corresponds to the areal entropy density, our E-MOTS reduces to the usual marginally outer trapped surface, i.e., the apparent horizon. Therefore, the E-MOTS can be regarded as an entropic generalization of the apparent horizon. We then showed that the key relation \eqref{ST = 1 -v del v SCv}, which connects the entropy evaluated on the E-MOTS $\mathcal{T}_{v}$ and that on a section of the background Killing horizon $\mathcal{C}_{v}$, holds at first-order. To apply this relation to the dynamical black hole entropy, we need to extend the definition of the black hole entropy from the background Killing horizon $\mathcal{H}^{+}$ to a neighborhood of it. We propose a possible extension given by Eq.~\eqref{swallhat}. Based on this extension, our relation \eqref{ST = 1 -v del v SCv} indicates that the Hollands--Wald--Zhang entropy evaluated on a section of the event horizon $\mc{C}_{v}$ is equivalent to the (extended) Wall entropy evaluated on the E-MOTS $\mc{T}_{v}$, as shown in Eq.~\eqref{main equation}. This generalizes the equivalence between the dynamical black hole entropy and the area of the apparent horizon in the case of the Einstein--Hilbert action, which is established in Refs.~\cite{Hollands:2024vbe, Visser:2024pwz}.
We note that our key relation \eqref{ST = 1 -v del v SCv} holds for any extension of the Wall entropy and our Eq.~\eqref{swallhat} is just an example.

We would like to emphasize the importance of our result~\eqref{main equation}, which demonstrates the robustness of the dynamical black hole entropy formula in the following sense. Let us consider a situation where a stationary black hole is perturbed by an injection of energy at a given moment, and subsequently settles down to another stationary black hole. During this process, the marginally outer trapped tube (MOTT), which is foliated by apparent horizons, lies on a null hypersurface in each of the stationary regimes before and after the energy injection, but deviates from null during the intermediate phase when energy is being injected. As a result, depending on the gauge choice, the MOTT in at least one stationary regime departs from the background Killing horizon $\mc{H}^{+}$. In contrast, when the entire spacetime is stationary, the well-established black hole entropy is evaluated on a Killing horizon which coincides with a MOTT. Thus, in order for the dynamical black hole entropy, which is evaluated on a single null hypersurface $\mc{H}^{+}$, to consistently represent the standard black hole entropy in each stationary phase, it is necessary to establish a relation between the entropy on $\mc{H}^{+}$ and on an apparent horizon. Our formula~\eqref{main equation} actually provides such a relation because an E-MOTS coincides with a usual marginally outer trapped surface, namely, an apparent horizon, in a stationary regime. Thus, our result ensures that, in each stationary regime, the dynamical entropy reduces to the Wall entropy on the apparent horizons, and which can be understood as the Iyer--Wald and the Wald entropy as well through Eq.~\eqref{entropy relation for stationary spacetime},
\begin{align}
S_{\text{HWZ}} = (S_{\text{Wall}})_{\text{app}} = (S_{\text{IW}})_{\text{app}} = S_{\text{Wald}} \qquad \text{in stationary regime}.
\end{align}
Note that the Wald entropy here is defined by considering an auxiliary, fully stationary spacetime that contains the stationary portion of the dynamical spacetime of interest.

This fact also opens up the possibility that discussions based on comparing the entropies of two stationary black hole spacetimes can be extended to the dynamical settings by employing the dynamical black hole entropy. In particular, generalizing arguments related to the weak cosmic censorship conjecture, which have been developed based on a comparison version of the second law for the Wald entropy (see Refs.~\cite{Lin:2022ndf,Wu:2024ucf,Yang:2025leh} and App.~B in Ref.~\cite{Yoshida:2024txh}), may be an interesting direction for future investigation.

Having established the properties of our formulation, we now turn to a comparison with other definitions of the entropic expansion that have been proposed in the literature.
 
As an extension of the expansion that incorporates entropy effects in the semi-classical regime, quantum expansion was proposed by Bousso, Fisher, Leichenauer, and Wall \cite{Bousso:2015mna}. 
For a section $\mathcal{C}$ of a given null hypersurface, let $v$ be an affine parameter of each generator and the surface $\mc{C}$ is parametrized as $v = V(\varphi^{A})$ by a function $V(\varphi^{A})$. The quantum expansion is then defined as  
\begin{align}
\Theta^{\text{BFLW}}(\varphi^{A}) \coloneqq  \frac{4 G \hbar}{\sqrt{\gamma_{{}_\mathcal{C}}}} \frac{\delta}{\delta V(\varphi^{A})} S_{\text{gen}}[\mathcal{C}]\,,
\end{align}
where $S_{\text{gen}}$ is the generalized entropy evaluated on a time slice $\Sigma_{\text{out}}$ with a boundary $\mathcal{C}$, which consists of the classical gravitational entropy and entanglement entropy of quantum fields with the interior of $\mathcal{C}$ traced out.
In the classical limit $\hbar \to 0$, the dominant contribution in the generalized entropy is expected to come from the gravitational part. When the gravitational entropy is expressed as the integral of an entropy density $s$ over $\mathcal{C}$, the expression by the functional derivative can be expressed by the Lie derivative along the generator and hence it reduces to
\begin{align}
\Theta^{\text{BFLW}} \stackrel{\hbar \to 0}{\approx} \frac{s}{s^{\text{Area}}} \Theta_{+}\,.
\label{classical limit of quantum expansion}
\end{align}
where $s^{\text{Area}}$ is the areal entropy density defined in Eq.~\eqref{sArea}. For the derivation of the expression \eqref{classical limit of quantum expansion} see, for example, App. B in Ref.~\cite{Kanai:2024zsw} for the similar calculations. Thus, the classical limit of the quantum expansion reduces to our entropic expansion up to normalization. Hence, our E-MOTS can be regarded as the MOTS with respect to the quantum expansion in the classical limit as well. It should be noted that although the quantum expansion is formulated for arbitrary null hypersurfaces, the choice of gravitational entropy to be used therein is not specified in Ref.~\cite{Bousso:2015mqa}. In the literature \cite{Fu:2017lps,Leichenauer:2017bmc, Kanai:2024zsw}, a naive application to the Dong--Camps entropy \cite{Dong:2013qoa, Camps:2013zua} is adopted for, for example, the Gauss--Bonnet gravity.  In contrast, we have proposed a possible extension of the Wall entropy based on the covariant phase space formalism, which may serve as a basis for a more consistent definition of the quantum expansion in future work. 

Finally, we note that the Wall entropy was originally formulated for first-order perturbations around stationary black holes. Extensions of this construction to higher orders have been explored within the framework of gravitational effective field theory, both at quadratic order \cite{Hollands:2022fkn} and to all perturbative orders within effective field theory \cite{Davies:2023qaa}.
In the present work, we have shown that the HWZ entropy on a horizon cross-section coincides with the Wall entropy evaluated on the corresponding E-MOTS. It would be interesting to investigate whether an analogous correspondence persists beyond the first-order analysis.

\section*{Acknowledgements}
H.F. is grateful to Robert Wald for insightful discussions on the dynamical black hole entropy during the 34th Midwest Relativity Meeting,  and also thanks Ratin Akhoury for hosting and supporting the visit to the University of Michigan, which enabled participation in the meeting and fruitful exchanges.
H.F. also thanks Atsuya Tokutake for helpful discussions on the covariant phase space formalism.
This work was partially supported by Grants-in-Aid for Scientific Research from the Japan Society for the Promotion of Science (JSPS) and the Ministry of Education, Culture, Sports, Science and Technology (MEXT) of Japan
under Grant Numbers
JP22H01217~(H.F.),
JP22J20380~(K.N.),  
JP21H05189~(D.Y.).
K.Y. is supported by JST SPRING, Grant Number JPMJSP2108 and by a research encouragement grant from the Iwanami F\=ujukai. 

\appendix

\section{A Procedure for Deriving Charges in Diffeomorphism Invariant Theories\label{app: Charges}}

As noted in Sec.~\ref {sec: review on HWZ}, the HWZ entropy is defined via the Noether charge approach in the covariant phase space formalism.
In this Appendix, we outline the procedure for deriving the building blocks of the HWZ entropy, such as $E_R^{abcd}$, $W^{abc}$, and others, from a diffeomorphism covariant Lagrangian without matter fields following the work of Iyer and Wald \cite{Iyer:1994ys}.
The Lagrangian $\mr{D}$-form is taken to be
\begin{align}
    \bm{L} &=\bm{\varepsilon}L(g_{ab},R_{abcd},\nabla_{a_1}R_{abcd},\nabla_{(a_1}\nabla_{a_2)}R_{abcd},\cdots, \nabla_{(a_1}\cdots \nabla_{a_m)}R_{abcd})\,.
\label{general gravitaional L without matter}
\end{align}
Here, $\bm{\epv}$ denotes the volume $\mr{D}$-form associated with $(\mc{M},g_{ab})$, and $\nabla_a$ and $R_{abcd}$ denote the covariant derivative and the Riemann curvature tensor associated with the metric $g_{ab}$, respectively.
The treatment of (bosonic) matter fields is similar, but we shall not consider them in this paper.

\subsection{Symplectic Potentials}

The first-order variation of the Lagrangian $\mr{D}$-form $\bm{L}$ can be expressed as
\begin{align}
    \delta \bm{L} = \bm{E}_g^{ab}(g) \delta g_{ab}+ d \bm{\theta}(g,\delta g)
    \label{formal form of variation of L}
\end{align}
with
\begin{align}
 \bm{E}_{g}^{ab}(g) = \bm{\varepsilon} E_{g}^{ab}(g)\,, 
\end{align}
where $E_g^{ab}(g) = 0$ denotes the equations of motion for the metric $g_{ab}$. We investigate the structure of these terms in the following.

Since the variation of the determinant of the metric $g$ is given by $\delta g = g g^{ab} \delta g_{ab}$, the variation of the volume form can be evaluated as 
\begin{align}
  \delta \bm{\varepsilon} = \bm{\varepsilon} \frac{1}{2} g^{ab} \delta g_{ab} \,,
\end{align}
then, the variation of the Lagrangian $\mr{D}$-form takes the form
\begin{align}
    \delta \bm{L}&= \bm{\epv} \biggl[\frac{1}{2}g^{ab} L\delta g_{ab} +\pdv{L}{g_{ab}} \delta g_{ab}  \nn
&+ \pdv{L}{R_{abcd}}\delta R_{abcd}+ \pdv{L}{\nabla_{b_1}R_{abcd}}\delta\left(\nabla_{b_1}R_{abcd} \right) +\cdots + \pdv{L}{\nabla_{(b_1}\cdots \nabla_{b_m)}R_{abcd}}\delta\left(\nabla_{(b_1} \cdots \nabla_{b_m)}R_{abcd} \right)
\biggr]\,.
\label{variation of L}
\end{align}
Here, $\pdv{L}{g_{ab}}$ denotes the variation of the Lagrangian with respect to the metric $g_{ab}$, treating $R$, $\nabla R$, and higher derivatives as independent of $g$. 
The same procedure applies to the definitions of $\pdv{L}{R_{abcd}}$, $\pdv{L}{\nabla_{b_1}R_{abcd}}$, etc.
To simplify the expression, we introduce the following shorthand notion:
First, let us express a collection of $n + 4$ abstract indices $a_{n+4} \cdots a_{4}a_{3}a_{2}a_{1}$ as $I_{n}$.
Then we can express the Riemann tensor as 
\begin{align}
&R_{I_0} = R_{a_4 a_3 a_2 a_1}\,.
\end{align}
Furthermore, let us introduce shorthand notions of the derivatives of the Riemann tensor as
\begin{align}
R_{a_{n+4} \cdots a_{1}} \coloneqq \nabla_{a_{n+4}} \cdots \nabla_{a_5} R_{a_{4}a_{3}a_{2}a_{1}}\,,
\end{align}
and by using the notion of a collection of the abstract indices, it can be expressed as 
\begin{align}
&R_{I_n} = R_{a_{n+4} a_{n+3} \cdots a_1}
\,.
\end{align}
Note that the relation
\begin{align}
R_{I_{n}} = \nabla_{a_{n+4}} R_{I_{n-1}}
\end{align}
is immediately followed from the definition.
Similarly, we introduce the notion
\begin{align}
&E_{n}^{I_n} \coloneqq \pdv{L}{\nabla_{(a_{n+4}} \cdots \nabla_{{a}_{5})}R_{a_4 a_3 a_2 a_1} }\, .
\end{align}
Using these notions, the variation of the Lagrangian can be expressed as 
\begin{align}
     \delta \bm{L}= \bm{\epv} \biggl[\frac{1}{2}g^{ab} L\delta g_{ab} +\pdv{L}{g_{ab}} \delta g_{ab} 
     + \sum_{n=0}^{m} E_{n}^{I_n} \delta R_{I_n} \biggr]\,.
\label{variation of L in our notation}   
\end{align}
In order to write $\delta \bm{L}$ in the form of Eq.~\eqref{formal form of variation of L}, we should decompose the last term into a total derivative part and a part proportional to the variation of the metric. In what follows, we first aim to rewrite the last term as
\begin{align}
    \sum_{n=0}^{m} E_{n}^{I_n} \delta R_{I_n} = \nabla_a V^a + A^{ab}\delta g_{ab} +E_R^{I_0}\delta R_{I_0}.
\end{align}
Since
\begin{align}
    \delta R_{I_n} &= \delta (\nabla_{a_{n+4}} R_{I_{n-1}}) \nn
    &= \nabla_{a_{n+4}} \delta R_{I_{n-1}} -\sum_{j=1}^{n+3} R_{a_{n+3}\cdots c \cdots a_1} \delta \Gamma^c_{a_{n+4}a_j}
    \nn &= \nabla_{a_{n+4}} \delta R_{I_{n-1}} 
    -\frac{1}{2}\sum_{j=1}^{n+3} \tens{R}{_{a_{n+3} \cdots} ^d _{\cdots a_1}} \left(\nabla_{a_{n+4}}\delta g_{da_j}+\nabla_{a_j}\delta g_{a_{n+4}d} -\nabla_d \delta g_{a_{n+4}a_j} \right),
\end{align}
applying the Leibniz rule $f\nabla g = \nabla(fg)- g\nabla f$, we obtain
\begin{align}
E_n^{I_n}\delta R_{I_n} &= \nabla_{a_{n+4}} \left[E_n^{a_{n+1}I_{n-1}}\delta R_{I_{n-1}}-S_n^{a_{n+4}bc}\delta g_{bc} \right]
    +\left[\nabla_{a_{n+4}}S_n^{a_{n+4}bc}\right]\delta g_{bc}  -\left[\nabla_{a_{n+4}} E_n^{a_{n+1}I_{n-1}}\right]\delta R_{I_{n-1}} ,
\end{align}
where
\begin{align}\notag
    S_n^{a_{n+4}bc} \coloneqq \frac{1}{2}\sum_{j=1}^{n+3}& \Big(
        \tens{R}{_{a_{n+3} \cdots } ^{(b|} _{\cdots a_1} }E_n^{a_{n+4}\cdots {|c)} \cdots a_1 } + \tens{R}{_{a_{n+3} \cdots } ^{(c|} _{\cdots a_1} }E_n^{|b) a_{n+3}\cdots a_{n+4}  \cdots a_1 }\\
        &
       - \tens{R}{_{a_{n+3} \cdots } ^{a_{n+4}} _{\cdots a_1} }E_n^{(b| a_{n+3}\cdots |c)  \cdots a_1 }
    \Big)\,.
\end{align}
Then, let us evaluate $\delta R_{I_{n-1}}$. By defining
\begin{align}
   (R_{n-1} \delta \Gamma)_{I_{n}}\coloneqq  \sum_{j=1}^{n+3} R_{a_{n+3}\cdots c \cdots a_1} \delta \Gamma^c_{a_{n+4}a_j}\,,
\end{align}
for $n \geq 1$, we obtain
\begin{align}
 \delta R_{I_{n}} = \delta \left( \nabla_{a_{n+4}} R_{I_{n-1}}\right) = \nabla_{a_{n+4}} \delta R_{I_{n-1}} - (R_{n-1} \delta \Gamma)_{I_n}.
\end{align}
Using this equation recursively, we obtain
\begin{align}
    \delta R_{I_n} = \nabla_{a_{n+4}}\cdots \nabla_{a_5} \delta R_{I_0}
    -\left[(R_{n-1}\delta \Gamma)_{I_n}+\nabla_{a_{n+4}}(R_{n-2}\delta \Gamma)_{I_{n-1}}+\cdots + \nabla_{a_{n+4}}\cdots \nabla_{a_6}(R_0 \delta \Gamma)_{I_1}\right]\,.
\end{align}
Since only the first term includes the variation of the Riemann tensor, by applying the Leibniz rule iteratively, we can express
\begin{align}
  &  E_n^{I_n}\delta R_{I_n} = \nabla_a \tilde{V}_n^a + \tilde{A}_n^{ab} \delta g_{ab} + (-1)^{n} \left[
    \nabla_{(b_1}\cdots \nabla_{b_n)} \pdv{L}{\nabla_{b_1}\cdots \nabla_{b_n} R_{abcd}}
    \right] \delta R_{abcd},\\
&\tilde{V}_n^a = \tilde{S}_n^{abc} \delta g_{bc} 
        + \sum_{i=0}^{n-1} T_{n,(i)}^{abcdeb_1\cdots b_i} \delta\left(\nabla_{(b_1}\cdots \nabla_{b_i)} R_{bcde}\right)\,,
\end{align}
where $\tilde{S}_n^{abc}$, $T_{n,(i)}^{abcdeb_1\cdots b_i}$, and $\tilde{A}_n^{ab}$ are tensor fields locally and covariantly constructed from the metric, the curvature, and their covariant derivatives.
We also note that $\tilde{S}_0 = T_{0,(i)} = \tilde{A}_0 = 0$.

We thus obtain the following expression for the variation of the Lagrangian:
\begin{align}
    \delta \bm{L}&= \bm{\epv} \biggl[
        \left( \frac{1}{2}g^{ab}L + \pdv{L}{g_{ab}} + \tilde{A}^{ab} \right)\delta g_{ab} + E_R^{abcd}\delta R_{abcd} \biggr] 
       \nn &\quad + \bm{\epv} \nabla_a \left[ \tilde{S}^{abc} \delta g_{bc} 
        + \sum_{i=0}^{m-1} T_{(i)}^{abcdeb_1\cdots b_i} \delta\left(\nabla_{(b_1}\cdots \nabla_{b_i)} R_{bcde}\right) \right]\,,
\end{align}
where
\begin{align}
    \tilde{S}^{abc} = \sum_{n=0}^{m} \tilde{S}_n^{abc}\,, \quad
    T_{(i)}^{abcdeb_1\cdots b_i} = \sum_{n=0}^{m} T_{n,(i)}^{abcdeb_1\cdots b_i}\,, \quad
    \tilde{A}^{ab} = \sum_{n=0}^{m} \tilde{A}_n^{ab}\,,
\end{align}
and
\begin{align}
    E_R^{abcd}= \pdv{L}{R_{abcd}} - \nabla_{b_1} \pdv{L}{\nabla_{b_1}R_{abcd}} + \cdots + (-1)^m \nabla_{(b_1}\cdots \nabla_{b_m)} \pdv{L}{\nabla_{b_1}\cdots \nabla_{b_m} R_{abcd}}\,.
\end{align}

Since $R_{abcd}$ depends on $g_{ab}$, its variation is given by
\begin{align}
    \delta R_{abcd} &= \tens{R}{_{abc}^e} \delta g_{ed} + 2g_{de} \nabla_{[b} \delta \Gamma^e_{a]c} 
    \nn &= \tens{R}{_{abc}^e} \delta g_{ed} + \nabla_{[b} \left(
        \nabla_{a]} \delta g_{dc} + \nabla_{|c|} \delta g_{a]d} - \nabla_{|d|} \delta g_{c|a]} \right)\,,
\end{align}
and hence we obtain
\begin{align}
    E_R^{abcd} \delta R_{abcd} &= E_R^{abcd} \tens{R}{_{abc}^e} \delta g_{ed}
    + \left(E_R^{adcb} + E_R^{bdac} - E_R^{cdba} \right) \nabla_d \nabla_a \delta g_{bc}\,.
\end{align}
Since $E_R^{abcd}$ possesses the same index symmetries as the Riemann tensor, the following equations hold:
\begin{align}
    E_{R}^{abcd} + E_{R}^{acdb} + E_{R}^{adbc} = 0\,, \qquad E_{R}^{abcd}=-E_{R}^{bacd}=-E_{R}^{abdc} = E_{R}^{cdab}\,.
    \label{symmetry of ER}
\end{align}
Hence, we can express
\begin{align}
    E_R^{abcd} \delta R_{abcd}&=  E_R^{abcd} \tens{R}{_{abc}^e}\delta g_{ed}+2E_R^{abcd}\nabla_d \nabla_a \delta g_{bc} 
    \nn&=\left[E_{R}^{b_1b_2b_3a}\tens{R}{_{b_1b_2b_3} ^b} + 2 \left(\nabla_{b_1}\nabla_{b_2}E_{R}^{b_1abb_2}\right)\right]\delta g_{ab}
\nn &\quad +2\nabla _a\left[ E_{R}^{ab_1b_2b_3} \nabla_{b_3}\delta g_{b_1b_2} - \left(\nabla_{b}E_{R}^{bb_1b_2a}\right)\delta g_{b_1b_2} \right]\,.
\end{align}

We thus finally arrive at
\begin{align}
    \delta \bm{L} = \bm{\epv} E_g^{ab} \delta g_{ab} + d \left( \bm{\epv}_a V^a \right)
\end{align}
with
\begin{align}
   & E_g^{ab} = \frac{1}{2}g^{ab}L + \pdv{L}{g_{ab}} + E_R^{b_1b_2b_3a} \tens{R}{_{b_1b_2b_3}^b} + A^{ab}\,, 
   \quad A^{ab} = \tilde{A}^{ab} + 2 \nabla_{b_1} \nabla_{b_2} E_R^{b_1abb_2}\,, \\
   & V^a = 2E_R^{abcd} \nabla_d \delta g_{bc} + S^{abc} \delta g_{bc} 
   + \sum_{i=0}^{m-1} T_{(i)}^{abcdeb_1\cdots b_i} \delta \left( \nabla_{(b_1} \cdots \nabla_{b_i)} R_{bcde} \right), 
   \quad S^{abc} = \tilde{S}^{abc} - 2 \nabla_d E_R^{dbca}\,.
\end{align}
Here, we have used the notation
\begin{align}
    (\bm{\epv}_a V^a)_{a_2\cdots a_\mr{D}} \coloneqq (\bm{\epv})_{aa_2\cdots a_\mr{D}} V^a = (\iota_V \bm{\epv})_{a_2\cdots a_\mr{D}}.
    \label{def epv_a app}
\end{align}

Comparing with Eq.~\eqref{formal form of variation of L}, we obtain 
\begin{align}
\bm{\theta}(g,\delta g) = \bm{\epv}_a \biggl[2{E}_R^{abcd}\nabla_d\delta g_{bc}+{S}^{abc}\delta g_{bc}+\sum_{i=0}^{m-1}{T}_{(i)}^{abcdeb_1\cdots b_i}\delta\bigl(\nabla_{(b_1}\cdots\nabla_{b_i)}R_{bcde}\bigr)
    \biggr]\,,
    \label{theta with no matter}
\end{align}
up to the ambiguity of an exact term $\bm{\theta} \to \bm{\theta} + d\bm{Y}$.

We also note that adding an exact form to the Lagrangian, $\bm{L} \to \bm{L} + d \bm{\lambda}$, does not affect the equations of motion, but it changes the boundary term as $\bm{\theta} \to \bm{\theta} + \delta \bm{\lambda}$.
These degrees of freedom coincide with the ambiguities discussed in Eq.~\eqref{ambiguities for theta}.

\subsection{Noether Charges\label{app: Noether Charges}}

The Noether current $(\mr{D}-1)$-form associated with a diffeomorphism generated by a vector field $\chi^a$ is given by
\begin{align}
    \bm{J}_\chi = \bm{\theta}(g,\lie{\chi} g) -\iota_\chi \bm{L}(g)\,.
    \label{def Noether current app}
\end{align}
Using the fact that $\mathsterling_\chi \bm{L} = d \iota_\chi \bm{L}$ and also $\mathsterling_\chi \bm{L} = \bm{E}_g^{ab} \lie{\chi} g_{ab} + d \bm{\theta}(g,\lie{\chi} g)$, we obtain\begin{align}
    d\bm{J}_\chi = -\bm{E}_g^{ab}(g) \lie{\chi} g_{ab} = - 2\bm{\epv} E_g^{ab} \nabla_{(a}\chi_{b)}\,.
    \label{d Noether current app}
\end{align}
We define the stress-energy-momentum tensor of the external matter fields by
\begin{align}
    T^{ab} \coloneqq -2 E_g^{ab}\,.
    \label{sem tensor}
\end{align}
Then, we find
\begin{align}
   d\bm{J}_\chi = d \left[ \bm{\epv}_a T^{ab} \chi_b \right] - \bm{\epv} \left( \nabla_a T^{ab} \right) \chi_b\,.
   \label{dJ in T}
\end{align}
Since the left-hand side of Eq.~\eqref{dJ in T} is an exact form, in order to satisfy this equation for arbitrary $\chi_{b}$, 
the following condition must hold:
\begin{align}
    \nabla_a T^{ab} = 0\,.
    \label{cavariant conservation of T}
\end{align}

Then, by the Poincar\'e's lemma, there exists a $(\mr{D}-2)$-form $\bm{Q}_\chi$ such that
\begin{align}
    \bm{J}_\chi - \bm{\epv}_a T^{ab} \chi_b = d \bm{Q}_\chi\,.
\end{align}
From this observation, the Noether current can be written as
\begin{align}
    \bm{J}_\chi = d \bm{Q}_\chi + \bm{C}_\chi\,,
    \label{def Q and C app}
\end{align}
where
\begin{align}
    \bm{C}_\chi = \bm{\epv}_a T^{ab} \chi_b = -2 \bm{\epv}_a E_g^{ab} \chi_b\,.
    \label{C expression without matter}
\end{align}
Note that $\bm{C}_\chi$ is linear in $\chi^a$ and vanishes under the equations of motion.

Let us investigate the structure of $\bm{Q}_\chi$.
From Eqs.~\eqref{theta with no matter} and \eqref{def Noether current app}, we find that $\bm{J}_\chi$ contains the following terms:
\begin{align}
    4\bm{\epv}_a E_R^{abcd} \nabla_d \nabla_{(b} \chi_{c)}  
    = \bm{\epv}_a \left[\nabla_d\left(-2 E_R^{adbc} \nabla_b \chi_c \right) + 2\left(\nabla_d E_R^{adbc}\right)\nabla_b \chi_c - 2 E_R^{acdb} \tens{R}{_{dbc}^e} \chi_e\right].
    \label{J chi ER term}
\end{align}
Here we use the identity
\begin{align}
E_{R}^{a(bc)d} = - \frac{1}{2} \left( E_{R}^{adbc} + 2 E_{R}^{acdb} \right),
\end{align}
which can be derived from Eq.~\eqref{symmetry of ER}.
Let us express the first term in the exact form. By introducing a notion
\begin{align}
    (\bm{\epv}_{ab} \beta^{ab})_{a_3 \cdots a_\mr{D}} \coloneqq (\bm{\epv})_{aba_3 \cdots a_\mr{D}} \beta^{ab}\,, 
    \label{def epv_ab app}
\end{align}
one can show the relation
\footnote{
Our notation can be expressed by the Hodge dual as follows, 
\begin{align}
 \bm{\epv}_{a} \alpha^{a} &= (-1)^{D-1} * \bm{\alpha}\,, \qquad \bm{\epv}_{ab} \beta^{ab} = 2 * \bm{\beta}\,,
\end{align}
where we use the convention of the Hodge dual such as
\begin{align}
(* \bm{\alpha})_{a_{1} \cdots a_{D-1}} = \epv_{a_{1} \cdots a_{D-1}}{}^{a_{D}} \alpha_{a_{D}}, \qquad (* \bm{\beta})_{a_{1} \cdots a_{D-2}} = \frac{1}{2} \epv_{a_{1}\cdots a_{D-2}}{}^{a_{D-1}a_{D}} \beta_{a_{D-1} a_{D}}\,.
\end{align}
Then, the relation \eqref{ep del beta = d ep beta} follows from the relation,
\begin{align}
(* d * \bm{\beta})_a = - \nabla_{d} \beta_{a}{}^{d}
\end{align}
with the property
$* * = (-1)^{p (D-p) + 1}$ valid when acting on $p$-form.
}
\begin{align}
\bm{\epv}_{a} \nabla_{d} \beta^{ad} = d \left( \frac{1}{2} \bm{\epv}_{ab} \beta^{ab} \right), \label{ep del beta = d ep beta}
\end{align}
for any anti-symmetric tensor $\beta^{ab}$.
Using this relation, the first term in Eq.~\eqref{J chi ER term} can be expressed as
\begin{align}
   \bm{\epv}_a \nabla_d\left(-2 E_R^{adbc} \nabla_b \chi_c \right) = d \bm{\alpha}_\chi\,,
   \quad \bm{\alpha}_\chi = -\bm{\epv}_{ab} E_R^{abcd} \nabla_{[c} \chi_{d]}\,.
   \label{Q second term}
\end{align}
This $\bm{\alpha}_\chi$ coincides with the second term in Eq.~\eqref{Q up to JKM}.

Noting that $\bm{J}_\chi$ can be decomposed into parts linear in $\nabla \chi$ and parts linear in $\chi$, we may write
\begin{align}
    \bm{J}_\chi = d\bm{\alpha}_\chi + \bm{\mu}_{(\text{linear in } \nabla \chi)} + \bm{\mu}_{(\text{linear in } \chi)}\,.
\end{align}
By iteratively applying the Leibniz rule, the term linear in $\nabla \chi$ can be written as
\begin{align}
    \bm{\mu}_{(\text{linear in } \nabla \chi)} = d(\bm{W}^a \chi_a) + \bm{\nu}_{(\text{linear in } \chi)}\,,
\end{align}
and thus the Noether current takes the form
\begin{align}
    \bm{J}_\chi = d\left[\bm{\alpha}_\chi + \bm{W}^a \chi_a \right] + (\bm{\mu} + \bm{\nu})_{(\text{linear in } \chi)}\,.
\end{align}

Recalling that the Noether current can always be written as $\bm{J}_\chi = d\bm{Q}_\chi + \bm{C}_\chi$, with $\bm{C}_\chi$ linear in $\chi^a$, we find that $(\bm{\mu} + \bm{\nu})_{(\text{linear in } \chi)} = \bm{C}_\chi$, and thus we can express the Noether charge as
\begin{align}
    \bm{Q}_\chi &= \bm{W}^a \chi_a + \bm{\alpha}_\chi = \bm{\epv}_{ab} \left[ W^{abc} \chi_c - E_R^{abcd} \nabla_{[c} \chi_{d]} \right]\,,
    \label{Q chi app}
\end{align}
up to an ambiguity of the form $\bm{Q}_\chi \to \bm{Q}_\chi + d \bm{Z}_\chi$.

By construction, $W^{abc}$ is a tensor field locally and covariantly constructed from the metric, the curvature tensors, and their covariant derivatives.
This is equivalent to the expression in Eq.~\eqref{Q up to JKM}.

Finally, since the ambiguities in $\bm{\theta} \to \bm{\theta} + \delta \bm{\lambda} + d\bm{Y}$ induce the corresponding ambiguities in $\bm{Q}_\chi$, we find that the Noether charge is ambiguous up to terms of the form
\begin{align}
 \bm{Q}_\chi \to \bm{Q}_\chi + \iota_\chi \bm{\lambda}(g) + \bm{Y}(g, \lie{\chi} g) + d \bm{Z}_\chi(g)\,.
 \label{Q ambiguities app}
\end{align}
This corresponds to Eq.~\eqref{Q ambiguities}, and these ambiguities are often referred to as the JKM ambiguities.

\section{Lie Derivatives of Scalar Densities}
\label{app:LieD}
In this section, we summarize the basic properties of a scalar density. A field $s(x)$ is said to be a scalar density of weight $p$ if it transforms under coordinate transformations $x \mapsto x'$ as
\begin{align}
s'(x') = \left| \det \frac{\partial x^{\mu}}{\partial x'{}^{\nu}} \right|^{p} s(x)\,.
\end{align}

The following properties immediately follows from the definition:
\begin{itemize}
\item A scalar density of weight $0$ is a scalar.
\item The weight of $s^{-1}$ is $ - p$ if $s$ has weight $p$. 
\item If $s_{1}$ and $s_{2}$ are scalar densities of weight $p_1$ and $p_2$, respectively, then their product $s_{1} s_{2}$ is a scalar density of weight $p_{1} + p_{2}$.
\end{itemize}

Let us consider infinitesimal coordinate transformation $x^{\mu} \to x'{}^{\mu} = x^{\mu} - t w^{\mu}$.
Then $s'(x')$ can be evaluated as
\begin{align}
s'(x - t w) &= \left|\det \left(\delta^{\mu}_{\nu} + t \frac{\partial w^{\mu}}{\partial x^{\nu}} \right)\right|^{p} s(x) \notag\\
&= (1 + t p \partial_{\mu} w^{\mu}) s(x) + \mathcal{O}(t^2)\,.
\end{align}
Then, the Lie derivative of a scalar density along a vector field $w^{a}$ is defined by
\begin{align}
\widetilde{\mathsterling}_{w} s (x) \coloneqq \lim_{t \to 0} \frac{s'(x) - s(x)}{t} = w^{\mu} \partial_{\mu} s(x) + p \partial_{\mu} w^{\mu} s(x)\,.
\end{align}
Then, $\widetilde{\mathsterling}_{w} s$ has the same weight as $s$.

In particular, the Lie derivative of a scalar density of weight-1 can be expressed as a total derivative, 
\begin{align}
\widetilde{\mathsterling}_{w} s = \partial_{\mu} (w^{\mu} s)\,.
\end{align}
Hence, our entropic expansion can be expressed by the Lie derivative
\begin{align}
\Theta_{w} = \frac{1}{s} \widetilde{\mathsterling}_{w} s\,.
\end{align}
From this expression, one can see that the entropic expansion is density-$0$, that is, a scalar under coordinate transformations.

\setstretch{0.65} 
\bibliography{ref} 
\bibliographystyle{utphys.bst}

\end{document}